\documentclass[12pt]{article}
\pdfoutput=1
\usepackage{amssymb,amsmath,slashed,latexsym}
\usepackage{xcolor}
\usepackage{tikz}
\usepackage[english]{babel}
\usetikzlibrary{calc}
\usetikzlibrary{matrix}
\usepackage[vcentermath]{youngtab}
\usepackage{float}
\usepackage{nicefrac}
\allowdisplaybreaks
   \usepackage[nosort]{cite}
\usepackage{color}
\newcommand{\ft}[2]{{\textstyle\frac{#1}{#2}}}

\def\rmi{{\rm i}}
\def\rmd{{\rm d}}
\newcommand{\hc}{{\rm h.c.}}
\newsavebox{\uuunit}
\sbox{\uuunit}
    {\setlength{\unitlength}{0.825em}
     \begin{picture}(0.6,0.7)
        \thinlines
        \put(0,0){\line(1,0){0.5}}
        \put(0.15,0){\line(0,1){0.7}}
        \put(0.35,0){\line(0,1){0.8}}
       \multiput(0.3,0.8)(-0.04,-0.02){12}{\rule{0.5pt}{0.5pt}}
     \end {picture}}

\csname @addtoreset\endcsname{equation}{section}
\newcommand{\SU}{\mathop{\rm SU}}
\newcommand{\SO}{\mathop{\rm SO}}
\newcommand{\U}{\mathop{\rm {}U}}
\newcommand{\USp}{\mathop{\rm {}USp}}

\newcommand{\su}{\mathfrak{su}}
   \usepackage[nosort]{cite}
   \pdfoutput=1
  \usepackage[pdftex]{hyperref}
\newcommand{\DSlr}{\stackrel{\leftrightarrow}{\DS}}
\newcommand{\susyvart}{\delta_{Q,S}(\susyparr{}{},\susyparrconformal{}{})}
\newcommand{\comm}[2]{[#1,#2]} 														
\newcommand{\acomm}[2]{\{#1,#2\}} 													
\newcommand{\di}{\slashed{\partial}}   												
\newcommand{\DS}{\slashed{D}}   														
\newcommand{\D}[2]{D_{#1}{#2}}   														
\newcommand{\der}[2]{\partial_{#1}^{#2}} 											
\newcommand{\derlr}[2]{\stackrel{\leftrightarrow}{\der{#1}{#2}}}					
\newcommand{\dilr}{\stackrel{\leftrightarrow}{\di}}									
\newcommand{\gammas}[2]{\gamma_{#1}^{#2}}											
\newcommand{\kron}[2]{\delta_{#1}^{#2}}												
\newcommand{\levi}[2]{\varepsilon_{#1}^{#2}}										
\newcommand{\La}{\mathcal{L}}  														
\newcommand{\s}{\enspace} 															
\newcommand{\curd}[2]{d_{#1}^{#2}} 													
\newcommand{\cure}[2]{e_{#1}^{#2}} 													
\newcommand{\curv}[2]{v_{#1}^{#2}} 													
\newcommand{\curlR}[0]{\lambda_R} 													
\newcommand{\curlL}[0]{\lambda_L} 													
\newcommand{\curxi}[2]{\xi_{#1}^{#2}} 												
\newcommand{\curt}[2]{t_{#1}^{#2}} 													
\newcommand{\curh}[2]{t_{#1}^{#2}} 													
\newcommand{\cursup}[2]{\mathcal{J}_{#1}^{#2}} 										
\newcommand{\curstr}[2]{\Theta_{#1}^{#2}} 											
\newcommand{\gad}[2]{D_{#1}^{#2}} 													
\newcommand{\gae}[2]{E_{#1}^{#2}} 													
\newcommand{\gav}[2]{V_{#1}^{#2}} 													
\newcommand{\gaa}[2]{\mathcal{A}_{#1}^{#2}} 													
\newcommand{\galR}[0]{\Lambda_R} 														
\newcommand{\galL}[0]{\Lambda_L} 														
\newcommand{\galRc}[0]{\overline{\Lambda}_R} 														
\newcommand{\galLc}[0]{\overline{\Lambda}_L} 														
\newcommand{\gaxi}[2]{\chi_{#1}^{#2}} 												
\newcommand{\gaxicc}[2]{\overline{\chi}_{#1}^{#2}} 									
\newcommand{\gat}[2]{T_{#1}^{#2}} 													
\newcommand{\gah}[2]{T_{#1}^{#2}} 													
\newcommand{\dilaton}[2]{b_{#1}^{#2}}												
\newcommand{\spcon}[2]{f_{#1}^{#2}}													

\numberwithin{equation}{section}
\newcommand{\gaz}[2]{\zeta_{#1}^{#2}} 													
\newcommand{\A}[2]{B_{#1}{#2}}														
\newcommand{\scalar}[2]{\varphi_{#1}^{#2}}												
\newcommand{\F}[2]{F_{#1}^{#2}}														
\newcommand{\spinor}[2]{\psi_{#1}^{#2}}												
\newcommand{\spinorc}[2]{\overline{\psi}{}_{#1}^{#2}}									
\newcommand{\vb}[2]{e_{#1}^{#2}}													
\newcommand{\gravitino}[2]{\psi_{#1}^{#2}}												
\newcommand{\gravitinoc}[2]{\overline{\psi}{}_{#1}^{#2}}									
\newcommand{\spgravitino}[2]{\phi_{#1}^{#2}}											
\newcommand{\spin}[2]{\omega_{#1}^{#2}}
\newcommand{\susyparr}[2]{\epsilon_{#1}^{#2}}										
\newcommand{\susyparrc}[2]{\overline{\epsilon}_{#1}^{#2}}							
\newcommand{\susyparrconformal}[2]{\eta_{#1}^{#2}}										
\newcommand{\susyparrconformalc}[2]{\overline{\eta}_{#1}^{#2}}							
\newcommand{\susycomm}{\comm{\delta(\susyparr{1}{})}{\delta(\susyparr{2}{})}}		
\newcommand{\susyvar}{\delta_Q(\susyparr{}{})}										
\newcommand{\susyvarspecial}{\delta_S(\eta)}

\newcommand{\imag}{\textrm{i}}
\newcommand{\N}{\mathcal{N}} 														

\usepackage[vcentermath]{youngtab}
\begin{document}

\begin{titlepage}
\begin{flushright}
arXiv:1702.06442
\end{flushright}
\vspace{.5cm}
\begin{center}
\baselineskip=16pt
{\LARGE    The $\N=3$ Weyl Multiplet in Four Dimensions  
}\\
\vfill
{\large Jesse van Muiden and Antoine Van Proeyen  
  } \\
\vfill
{\small  KU Leuven, Institute for Theoretical Physics,\\
       Celestijnenlaan 200D B-3001 Leuven, Belgium.
      \\[2mm] }
\end{center}
\vfill
\begin{center}
{\bf Abstract}
\end{center}
{\small The main ingredient for local superconformal methods is the multiplet of gauge fields: the Weyl multiplet. We construct the transformations of this multiplet for ${\cal N}=3$, $D=4$. The construction is based on a supersymmetry truncation from the $\N=4$ Weyl multiplet, on coupling with a current multiplet, and on the implementation of a soft algebra at the nonlinear level, extending $\su(2,2|3)$. This is the first step towards a superconformal calculus for ${\cal N}=3$, $D=4$.}\vspace{2mm} \vfill \hrule width 3.cm
{\footnotesize \noindent e-mails: jesse.vanmuiden@kuleuven.be,
antoine.vanproeyen@fys.kuleuven.be }
\end{titlepage}
\addtocounter{page}{1}
 \tableofcontents{}
\section{Introduction}
In supergravity it is often a non trivial exercise to construct theories in which gravity multiplets are coupled to matter multiplets, so called matter-coupled supergravity theories. The most systematic approach that has been used for such constructions is the superconformal method. In this method one starts with a supergravity theory that has additional conformal symmetries. The multiplet of gauge fields in these theories is called the Weyl multiplet. This forms the background for general superconformal-invariant interactions of matter multiplets. The superconformal symmetry is then broken to super-Poincar\'{e} symmetry
\cite{Kaku:1978ea}. This procedure is a far easier task than constructing a Poincar\'e supergravity from scratch. Furthermore, the extra symmetries in conformal supergravity also offer a systematic approach to the construction of the matter couplings and reveal much of its structure. This superconformal method has therefore been used extensively in the past for different supergravity theories in multiple dimensions and number of supersymmetries. The Weyl multiplet is also the basis for constructing conformal supergravity theories, which are theories with higher derivatives. E.g. recently the ${\cal N}=4$ conformal supergravity theories have been studied in \cite{Ciceri:2015qpa,Butter:2016mtk}.

An old argument says that ${\cal N}=3$ rigid supersymmetry theories always have a fourth supersymmetry, and hence are in fact ${\cal N}=4$ supersymmetric theories. This is based on theories that have an action with a $\sigma$-model for the scalars. For every supersymmetry added to the $\N=1$ case one needs an additional parallel complex structure, see \cite{Alvarez-Gaume:1981hm,Bagger:1984ge} for a more detailed discussion on this fact. Consequently, for $\N=3$ theories one needs two of such structures. However, similarly as for the complex numbers, two complex structures automatically lead to a third one. This then immediately results in a theory that has four supersymmetries. Another argument for the non-existence of $\,{\cal N}=3$ supersymmetric theories that are not ${\cal N}=4$ theories comes from the analysis of multiplets of massless states. The irreducible representations with maximal spin 1 carry either helicities $(1, 3\times 1/2, 3\times 0, -1/2)$ or $(1/2, 3\times 0, 3\times -1/2, -1)$. Both separately are not CPT invariant. To ensure CPT invariance, one should join both representations. The field content is then that of an ${\cal N}=4$ theory. The kinetic action that one can construct with at most 2 spacetime derivatives shows that indeed the fourth supersymmetry is always present. Recently, however, a completely new approach has been taken to look at four-dimensional $\N=3$ theories, which has taken quite some interest from the community \cite{Aharony:2015oyb,Garcia-Etxebarria:2015wns,Nishinaka:2016hbw,Aharony:2016kai,Imamura:2016abe,Agarwal:2016rvx,Garcia-Etxebarria:2016erx,Lemos:2016xke}. These approaches do not rely on the usual Lagrangian philosophy and use techniques originating from holography to discover new physics in these theories.

This inclusion of ${\cal N}=3$ in ${\cal N}=4$ does not hold for supergravity. In that case other representations, going up to spin~2 distinguish ${\cal N}=3$ from ${\cal N}=4$. The structure of the ${\cal N}=3$ and ${\cal N}=4$ supercurrents and their relation to ${\cal N}=6$ supergravity in 5 dimensions was considered in \cite{Ferrara:1998zt}. A superspace approach to conformal supergravity for ${\cal N}\leq 4$ has been given in \cite{Howe:1980sy}.
Poincar\'e supergravity-matter couplings with $\N=3$ in four dimensions have been considered for their representation content \cite{Rivelles:1981qy}, in (harmonic) superspace \cite{Brink:1978sz,Galperin:1986id} and using the group-manifold approach \cite{Castellani:1985ka}.\footnote{The $\N=3$ Poincar\'e supergravity theory in four dimensions was also studied in terms of the super-BEH effect in \cite{Ferrara:1985ft}.} However, it was never done using the superconformal method. It can be expected that these methods will result in the same theory for the supergravity obtained by promoting 3 of the 4 rigid supersymmetries in the ${\cal N}=4$ supersymmetric $\sigma$-model. In any case the superconformal construction of $\N=3$ Poincar\'e supergravity theory may shed more light on the structure of these theories whose solutions were recently studied in the context the AdS/CFT correspondence \cite{Karndumri:2016miq,Karndumri:2016tpf}. Also the alternative ${\cal N}=3$ theories mentioned above may be studied with superconformal methods.

The first step in this programme is constructing the Weyl multiplet. In fact, a separate motivation for the construction of the Weyl multiplet is that it seems to be the final gap in the construction of gauge multiplets of conformal supergravities. These theories have been constructed for all the possible superconformal groups in dimensions varying from $D=2$ to $D=6$.
It is proven using an algebra argument by Nahm in \cite{Nahm:1978tg} that no superconformal algebra exists in higher dimensions.\footnote{Note, however, that a Weyl multiplet has been constructed for $D=10$ in \cite{Bergshoeff:1983az}, which is not based on a linear superalgebra.} Much work has been done to find the Weyl multiplets in different dimensions and with a different number of supersymmetries, see e.g. \cite{Salam:1989fm}. In two dimensions the possibility for Weyl multiplets has been discussed in \cite{Bergshoeff:1985gc,McCabe:1986vb}. In three dimensions the different multiplets are discussed in \cite{vanNieuwenhuizen:1985cx,Rocek:1985bk,Lindstrom:1989eg}. An off shell formulation of the possible Weyl multiplets in three dimensions was given in \cite{Butter:2013goa,Butter:2013rba}. In four dimensions the Weyl multiplet for ${\cal N}=1$ has already been found in  \cite{Kaku:1978nz}. For ${\cal N}=2$ it was found in  \cite{deWit:1980ug} and for ${\cal N}=4$ in \cite{Bergshoeff:1980is}. It was shown in \cite{deWit:1978pd} that no Weyl multiplets exist beyond ${\cal N}=4$. For five dimensions only ${\cal N}=2$ appears in Nahm's classification, and the corresponding Weyl multiplets were found in \cite{Kugo:2000hn,Bergshoeff:2001hc}. The Weyl multiplets in 6 dimensions were constructed for $(1,0)$ supersymmetry in \cite{Bergshoeff:1986mz} and for $(2,0)$ in \cite{Bergshoeff:1999db} and these are the only cases that appear in the classification of Nahm. We would like to note that the given list of papers discussing the construction of Weyl multiplets and conformal supergravity theories is far from complete. Very interesting work on the subject has been done since the original multiplets were constructed. However, the goal of the given list is merely to present a short overview of the original works concerning the Weyl multiplets in different dimensions. For the case of $\N=3$ conformal supergravity in four dimensions suggestions have been made for the field content of the Weyl multiplet \cite{Fradkin:1985am,vanNieuwenhuizen:1985dp}. The actual derivation of the full symmetry transformations, however, has never been done.

Before discussing the explicit content of this paper we will give a small introduction into some general features of the Weyl multiplets in four dimensions. The Weyl multiplet is a massive multiplet with maximal spin~2. For every spin, the fields of massive multiplets form representations of $\USp(2{\cal N})$ \cite{Ferrara:1980bh,Ferrara:1980ra}. These are given in Table \ref{tbl:USP2Nreps}. The explicit $\SO(3,1)\times \SU({\cal N})$ representations known in four dimensions will be discussed in Section \ref{SSec:Components of the Weyl multiplet(s)} in more detail.
\begin{table}[htbp]
\begin{center}
$\begin{array}{|r|ccccc|}\hline
  J & 2 & \nicefrac32 & 1 & \nicefrac12 & 0 \\ \hline
  {\cal N}=1 & 1 & 2 & 1 &   &   \\
  {\cal N}=2 & 1 & 4 & 5+1 & 4 & 1 \\
  {\cal N}=3 & 1 & 6 & 14+1 & 14'+6 & 14 \\
  {\cal N}=4 & 1 & 8 & 27 & 48 & 42\\ \hline
\end{array}$
\end{center}
  \caption{\it $\USp(2{\cal N})$ representations of spin $J$ content of Weyl multiplets. }\label{tbl:USP2Nreps}
\end{table}

We have constructed the Weyl multiplet for $\N = 3$ conformal supergravity and found the nonlinear supersymmetry transformations of the fields in the multiplet. The applied method resulted in one Weyl multiplet. However, in other dimensions and with a different amount of supersymmetries it was found that there are several Weyl multiplets. Future work will have to determine if this is also the case for the $\N=3$ theory in four dimensions. The constructed Weyl multiplet consists of $64+64$ fermionic and bosonic components. The supersymmetric algebra applied on this representation resulted in a consistent soft algebra as is usually the case for gauged supersymmetric theories.

Previous research in the subject has often made use of the so-called method of current multiplets \cite{Ogievetsky:1976qc,Bergshoeff:1980is,Howe:1981qj}. This method requires a rigid supersymmetry theory at the outset. Because there is no known $\,\N = 3$ rigid supersymmetric field theory we have applied a three-step procedure. The first step in this method consists of consistently truncating the $\N = 4$ current multiplet of the ${\cal N}=4$ Maxwell multiplet, obtained in \cite{Bergshoeff:1980is}, to three supersymmetries. In the second step, similar to the case of $\N = 4$ conformal supergravity, the Weyl multiplet is found by coupling a field
to every current. By imposing invariance of the first order action (which consists of the $current \times field$ terms) one is able to derive the linear supersymmetry variations of the fields in the Weyl multiplet, starting from the variations of the currents.
These transformations have then been checked for consistency with the symmetry algebra $\su( 2, 2 | 3 )$. In the third step, the nonlinear supersymmetry variations are found by taking a general ansatz for the nonlinear terms and checking for consistency with the soft algebra, the chiral weights and the conformal weights of the fields.
\medskip

This paper is organized as follows. In Section \ref{Sec:The N=4 Weyl Multiplet} we recapitulate the already known Weyl multiplet for $\N=4$ conformal supergravity and its construction using the current method.

The truncation to $\N=3$ conformal supergravity will be discussed in Section \ref{Sec:The N=3 Weyl multiplet}. This will lead to the representation of the gravity multiplet (Weyl multiplet) for $\mathfrak{su}( 2, 2 | 3 )$. The $R$-symmetry group will break from $\SU(4)$ into $\U(3)$. The truncated Weyl multiplet, as already mentioned, consists of $64+64$ independent components while the $\N=4$ multiplet consists of $128+128$ independent components. A comparison of all the known Weyl multiplets in four dimensions is given in Section \ref{SSec:Components of the Weyl multiplet(s)}.

In Section \ref{SSec:The gauge fields and their transformations} we will give the explicit linear and nonlinear supersymmetry variations of the $\N=3$ Weyl multiplet and discuss the method used to find them. The nonlinear variations are discussed in detail in \ref{SSec:The nonlinear supersymmetry transformations}. An important part of this derivation leads to the structure of the soft algebra of the theory, which is a modification of the algebra $\su(2,2|3)$. Such soft algebras contain structure functions depending on the usual infinitesimal parameters as well as on the fields present in the Weyl multiplet itself.

Finally, in Section \ref{ss:conclusion} a discussion of the found results will be given.

Appendix \ref{app: Usefull identities and calculations} discusses some of the conventions and identities that have been used throughout the paper.

\section{The \texorpdfstring{$\N = 4$}{N=4} Weyl Multiplet and the current method}\label{Sec:The N=4 Weyl Multiplet}

The content of the $\N=4$ Weyl multiplet was given in \cite{Howe:1980du,Bergshoeff:1980sw,Siegel:1980bp} and the transformation rules were fully constructed in \cite{Bergshoeff:1980is,Bergshoeff:1983qk}. The method used was the so called supercurrent method. The philosophy is to start with a rigid supersymmetric theory and perturb the theory
around flat space. In this sense one splits the Lagrangian of the theory in a zeroth order and a first order part:
\begin{align}\label{Eq: Lagrangian expansion}
\La = \La_0  + \La_1 + \ldots .
\end{align}
The zeroth order Lagrangian contains the kinetic terms of the rigid supersymmetry theory and the first order part contains the coupling of the currents with the fields in the multiplet. This first order coupling uniquely determines as well the content of the Weyl multiplet as its linear supersymmetry variations. We will recapitulate this construction of \cite{Bergshoeff:1980is} in Section~\ref{SSec:The N=4 Weyl multiplet and its linear variations}. The nonlinear variations are determined by consistency with the appropriate algebra. This is discussed in Section \ref{SSec:The nonlinear variations of the N=4 Weyl multiplet}.

The rigid supersymmetric theory from which the $\N=4$ Weyl multiplet in four dimensions was constructed is the unique $\N=4$ SYM (super-Yang--Mills) theory that was introduced in \cite{Brink:1976bc}. Section \ref{SSec:The rigid N=4 SYM multiplet} will summarize the field content of this theory.

\subsection{The rigid \texorpdfstring{$\N=4$}{N=4} SYM multiplet}\label{SSec:The rigid N=4 SYM multiplet}

The rigid ${\cal N}=4$ super-Maxwell theory contains a gauge field $\A{\mu}{}$, which is an $\SU(4)$ singlet, four fermions $\spinor{i}{}$, in the $\SU(4)$ vector representation and six scalars, $\scalar{ij}{}=-\scalar{ji}{}$, combined in the $\mathbf{6}$ representation of $\SU ( 4 )$. The Latin indices denote the representation of the fields with respect to this $\SU(4)$ $R$-symmetry group and run from 1 to 4. We use chiral spinors $\psi ^i=P_L\psi ^i$ and their complex (or charge) conjugates $\psi_i=P_R\psi_i$. We use the notations from \cite{Freedman:2012zz} to indicate the chiralities.
The scalar fields obey the following reality relations
\begin{align}
\scalar{}{ij} = \scalar{ij}{*} = -\frac{1}{2}\levi{}{ijk\ell}\scalar{k\ell}{}.
\end{align}
Instead of using the vector $\A{\mu}{}$ we will often rather use its field strength, or even better the (anti-)selfdual parts of this tensor:
\begin{align}
\F{\mu \nu }{} =  2\partial _{[\mu }\A{\nu ]}{}, \qquad \F{\mu \nu }{\pm} = \frac{1}{2}( \F{\mu \nu }{} \pm \tilde{F}_{\mu \nu }).
\end{align}
The Lagrangian is given by
\begin{align}
\La = -\frac{1}{4} \F{\mu\nu}{} \F{}{\mu\nu}   - \frac{1}{2} \spinorc{}{i} \dilr \spinor{i}{} - \frac{1}{2} (\der{}{\mu} \scalar{ij}{})(\der{\mu}{} \scalar{}{ij}),
\label{LagrN4SYM}
\end{align}
which is conformally invariant, using the coordinate transformations with
\begin{equation}
\xi^\mu(x)=a^\mu +\lambda ^{\mu\nu}x_\nu+\lambda_{\rm D} x^\mu
+(x^2\lambda_{\rm K}^\mu-2x^\mu x^\nu  \lambda_{{\rm K}\nu }) \,, \label{ximu}
\end{equation}
and intrinsic dilatation transformations\footnote{See \cite[Sec. 15.3]{Freedman:2012zz} for more details.}
\begin{equation}
\delta _{\rm D}B_\mu =0\,,\qquad \delta _{\rm D}\scalar{ij}{}=\lambda _{\rm D}\scalar{ij}{}\,,\qquad \delta _{\rm D}\psi ^i=\ft32 \lambda_D \psi ^i\,.
\label{extracondphipsi}
\end{equation}
On top of that, it is invariant for the following supersymmetric variations ($\epsilon ^i=P_L\epsilon ^i$ and $\epsilon _i=P_R\epsilon _i$)
\begin{equation}
\begin{aligned}
\susyvart \A{\mu}{} &= \frac{1}{2}\susyparrc{}{i} \gammas{\mu}{} \spinor{i}{}  + \hc\,, \\
\susyvart \psi^i &= -\frac{1}{4} \gammas{}{\mu\nu}  \F{\mu\nu}{-} \susyparr{}{i} -  \di \scalar{}{ij} \susyparr{j}{}-2\scalar{}{ij}\eta_j\,, \label{Eq:SUSY variations of N=4 SYM}\\
\susyvart \scalar{ij}{} &= \susyparrc{[i}{} \spinor{j]}{} - \frac{1}{2} \levi{ijk\ell}{} \susyparrc{}{k} \spinor{}{\ell}.
\end{aligned}
\end{equation}
The action is invariant under rigid superconformal transformations, which means that $\epsilon ^i$ can be linear in the spacetime variable $x^\mu $:
\begin{equation}
  \epsilon ^i(x)= \epsilon ^i_0 + x^\mu \gamma _\mu \eta ^i\,,
 \label{epsilonineta}
\end{equation}
where $\epsilon ^i_0$ and $\eta ^i=P_R\eta ^i$ are constants.

The superconformal transformations satisfy the superconformal ${\cal N}=4$ algebra, which includes $\su(4)$ R-transformations
\begin{equation}
  \delta _{\rm U}\varphi _{ij}= 2\lambda _{[i}{}^k\varphi _{j]k}\,,\qquad \delta _{\rm U}\psi _i=- \lambda _{i}{}^j\psi _j\,,
 \label{delUlambda}
\end{equation}
with an anti-hermitian traceless parameter $ \lambda _{i}{}^j$. These transformations satisfy the algebra $\su(2,2\vert  4)$. The general form of the superconformal algebra $\su(2,2\vert \N)$ has been found in \cite{Ferrara:1977ij} for general $\,{\cal N}$. It is generated by the standard conformal generators of $\mathfrak{su}(2,2)$, together with supersymmetry generators ($Q^i=P_RQ^i$ and $Q_i=P_LQ_i$), special supersymmetry generators ($S^i=P_LS^i$ and $S_i=P_RS_i$) and the $\mathfrak{(s)u}(\N)$ $R$-symmetry generators $U_i{}^j$ (and $T$).
The non-trivial commutation relations are as follows:
\begin{equation}\label{Eq: Commutators of superconformal group}
\begin{aligned}[c]
\comm{M_{[ab]}}{M_{[cd]}} &= 4 \eta_{[a[c}M_{[d]b]]},\\
[K_{a}, M_{bc}] &= 2\eta_{a[b}K_{c]},\\
[P_{a},K_{b}] &= 2(\eta_{a b}D + M_{ab}),\\
\comm{M_{ab}}{Q^i_{\alpha}} &= -\frac{1}{2} (\gammas{ab}{}Q^i)_{\alpha}, \\
\comm{D}{Q_{\alpha}^i} &= \frac{1}{2}Q_{\alpha}^i, \\
\comm{U_{i}^{\s j}}{Q_{\alpha}^k} &= \kron{i}{k} Q_{\alpha}^j - \frac{1}{{\cal N}} \kron{i}{j} Q_{\alpha}^k, \\
\comm{U^{\s j}_i}{Q_{\alpha k}} &= -\kron{k}{j} Q_{\alpha i} + \frac{1}{{\cal N}} \kron{i}{j} Q_{\alpha k}, \\
\comm{T}{Q^i_{\alpha}} &= \frac{1}{2} \imag  Q^i_{\alpha}, \\
\comm{K_a}{Q^i_{\alpha}} &= (\gammas{a}{}S^i)_{\alpha}, \\
\acomm{Q_{\alpha i}}{Q^{\beta j}} &= - \frac{1}{2} \kron{i}{j} (\gammas{}{a})_{\alpha}^{\s \beta} P_a ,
\end{aligned}
\qquad
\begin{aligned}[c]
\comm{P_a}{M_{[bc]}} &= 2\eta_{a[b}P_{c]}, \\
[D,P_{a}] &= P_{a},\\
[D,K_{a}]&=-K_{a},\\
\comm{M_{ab}}{S^i_{\alpha}} &= -\frac{1}{2} (\gammas{ab}{}S^i)_{\alpha}, \\
\comm{D}{S_{\alpha}^i} &= -\frac{1}{2}S_{\alpha}^i ,\\
\comm{U_{i}^{\s j}}{S_{\alpha}^k} &= \kron{i}{k} S_{\alpha}^j - \frac{1}{{\cal N}} \kron{i}{j} S_{\alpha}^k, \\
\comm{U^{\s j}_i}{S_{\alpha k}} &= -\kron{k}{j} S_{\alpha i} + \frac{1}{{\cal N}} \kron{i}{j} S_{\alpha k}, \\
\comm{T}{S^i_{\alpha}} &= -\frac{1}{2} \imag  S^i_{\alpha}, \\
\comm{P_a}{S^i_{\alpha}} &= (\gammas{a}{}Q^i)_{\alpha}, \\
\acomm{S_{\alpha i}}{S^{\beta j}} &= - \frac{1}{2} \kron{i}{j} (\gammas{}{a})_{\alpha}^{\s \beta} K_a ,
\end{aligned}
\end{equation}
\begin{equation*}
\acomm{Q_{i\alpha}}{S^{j\beta}} = - \frac{1}{2}\kron{i}{j}\kron{\alpha}{\beta} D - \frac{1}{4} \kron{i}{j} (\gammas{}{ab})^{\s\beta}_{\alpha} M_{ab} +\frac{4-\mathcal{N} }{2{\cal N}}\rmi \kron{i}{j} \kron{\alpha}{\beta} T - \kron{\alpha}{\beta} U_{i}^{\s j}.
\end{equation*}
Remark that the $\U(1)$ generator $T$ is optional\footnote{One may distinguish in this way P$\SU(2,2|4)$ from $\SU(2,2|4)$, where the former does not contain the $\U(1)$ and defines a simple superalgebra. On the other hand, in  \cite{Intriligator:1998ig} this algebra is indicated as $\mathop{\rm PU}(2,2|4)$ and the $T$ symmetry is indicated there as $\U(1)_Y$.} for ${\cal N}=4$. It does not appear at the right-hand side of  (\ref{Eq: Commutators of superconformal group}). This implies that representations of the ${\cal N}=4$ superconformal algebra may or may not have such chiral transformations. One can check that the supersymmetry variations of the SYM multiplet in (\ref{Eq:SUSY variations of N=4 SYM}) constrain the chiral weights of $\susyparr{i}{}$ and $\susyparr{}{i}$ in that representation to be zero. However, if we omit $B_\mu $ and take instead $F^\pm_{\mu \nu }$ as independent field, then consistent chiral weights can be assigned. These chiral transformations \cite{Bergshoeff:1980is,Ferrara:1998zt,Intriligator:1998ig} will play an important role in this paper. The Weyl multiplet of ${\cal N}=4$ is a representation of $\SU(2,2|4)$, i.e. the fields do
transform under the $T$ symmetry. We will leave this aside for the moment, and will come back to this symmetry at the end of Sec. \ref{SSec:The N=4 Weyl multiplet and its linear variations}.

The translations are realized on the fields of the super-Maxwell multiplet as derivatives $P_\mu =\partial _\mu $, except on the gauge field, where they act as covariant translations: $P_\mu B_\nu= F_{\mu \nu }$.
The generators are related to transformations with parameters as
\begin{equation}
 \delta =\xi ^aP_a+ \ft12 \lambda ^{ab}M_{[ab]}+\lambda_{\rm D} D +
\lambda_{\rm K}^a K_a+  \lambda _i{}^jU_j{}^i+ \lambda_T\,T+\,\bar \epsilon ^i Q_i +\bar \epsilon _iQ^i+\bar \eta  ^i S_i +\bar \eta  _iS^i  \,.
 \label{parameters}
\end{equation}
In this way, the last anticommutator in \eqref{Eq: Commutators of superconformal group} e.g. implies
\begin{equation}
\begin{aligned}
  \left[\delta _S(\eta ),\delta _Q(\epsilon )\right]=& \delta _{\rm D}(\lambda_{\rm D}) + \delta _{\rm M}(\lambda ^{ab}) + \delta _{\rm U}(\lambda _i{}^j) + \delta_T(\lambda_T)\,, \\
  \lambda_{\rm D}=&\ft12(\epsilon ^i\eta _i+\hc)\,,\\
  \lambda ^{ab}=& \ft12(\epsilon ^i\gamma ^{ab}\eta _i+\hc)\,,\\
  \lambda _i{}^j=&\bar \epsilon ^j\eta _i-\bar \epsilon _i\eta ^j-\frac1{\cal N}\delta _i^j(\bar \epsilon^k\eta _k-\bar \epsilon _k\eta ^k)\,, \\
  	\lambda_{T} =& \ft{{\cal N}-4}{2{\cal N}} \imag \susyparrc{}{i} \susyparrconformal{i}{} + \hc.
\end{aligned}\label{thirdcoeffQS}
\end{equation}
The superconformal symmetries are rigid symmetries, but the $\U(1)$ gauge transformation of $B_\mu $ is local. Similar to other supersymmetric gauge theories, one then finds soft algebras. This means that the commutator relations of ($Q$ and $S$) symmetries lead to $\U(1)$ transformations with parameters dependent on the fields inside the multiplet itself.\footnote{In superspace this effect is due to the fact that we have taken a Wess-Zumino gauge, and these transformations are the decomposition rules to stay in this gauge.}
One finds that the supersymmetry commutator acting on the gauge field $\A{\mu}{}$ gives the following result:
\begin{equation}
\begin{aligned}
\comm{\delta_{Q,S}(\epsilon_1(x),\eta_1)}{\delta_{Q,S}(\epsilon_2(x),\eta_2)}\A{\mu}{} &= -\frac{1}{2} (\susyparrc{1}{i}(x)\gammas{}{\nu}\susyparr{2i}{}(x) + \hc)F_{\nu\mu} + \der{\mu}{} \lambda_{\U(1)}
\end{aligned}
\end{equation}
where $\lambda_{\U(1)}$ is the gauge parameter and the $x$-dependence of $\susyparr{}{}(x)$ is given in \eqref{epsilonineta}. This gauge parameter is now dependent on the fields in the multiplet, explicitly one has that
\begin{equation}
\lambda_{\U(1)} = -  \susyparrc{2}{i}(x) \susyparr{1}{j}(x) \scalar{ij}{}.
\end{equation}
One can check that the supersymmetry variations given in equation \eqref{Eq:SUSY variations of N=4 SYM} are consistent with the algebra given in \eqref{Eq: Commutators of superconformal group} plus the mentioned gauge parameter term.

Using the Noether procedure, the stress energy tensor, supercurrent and $R$-symmetry currents are found from the Lagrangian:
\begin{equation}
\begin{aligned}
\curstr{\mu\nu}{} &= -4 \F{\mu\lambda}{+} \F{\nu}{-\lambda} - \spinorc{}{i} \gammas{(\mu}{}\derlr{\nu)}{}\spinor{i}{} + \eta_{\mu \nu} (\der{}{\rho}\scalar{ij}{})(\der{\rho}{}\scalar{}{ij}) \\
&\s \s -2(\der{\mu}{}\scalar{ij}{})(\der{\nu}{}\scalar{}{ij}) - \frac{1}{3}(\eta_{\mu\nu} \square - \der{\mu}{} \der{\nu}{} )(\scalar{ij}{} \scalar{}{ij})\,, \\
\cursup{ \mu i}{} &=  -\frac{1}{2} \gammas{}{\nu\rho} \F{\nu\rho}{-} \gammas{\mu}{} \spinor{i}{} - 2\scalar{ij}{}\derlr{\mu}{} \spinor{}{j} - \frac{2}{3} \gammas{\mu\lambda}{} \der{}{\lambda} (\scalar{ij}{} \spinor{}{j})\,, \\
\curv{\mu j}{\s \;i} &= \scalar{}{ik} \derlr{\mu}{} \scalar{kj}{}  + \spinorc{}{i} \gammas{\mu}{} \spinor{j}{}  -  \frac{1}{4} \kron{j}{i} \spinorc{}{k}\gammas{\mu}{} \spinor{k}{}\,.
\end{aligned}
\label{currentsN4}
\end{equation}
These currents are, as usual, only determined modulo equations of motion of the matter fields in the SYM multiplet. These equations of motion are given by
\begin{align}
\partial ^\mu F_{\mu \nu }=0, \qquad \di \spinor{}{i} = 0, \qquad \square \scalar{ij}{}=0.
\label{eomN4}
\end{align}
The stress energy tensor is improved such that it is symmetric and traceless.\footnote{The tracelessness reflects the conformal invariance.} Furthermore, current conservation and conformal symmetry tells us that
\begin{align}
\der{}{\mu}\curstr{\mu\nu}{} = 0, \qquad \der{}{\mu} \curv{\mu j}{\s \;i}= 0, \qquad \der{}{\mu} \cursup{\mu i}{} = 0, \qquad \eta ^{\mu \nu }\curstr{\mu\nu}{} = 0\qquad \gammas{}{\mu} \cursup{\mu i}{} = 0.
\label{curvatureconstraints}
\end{align}
The last two equalities are a consequence of the conformal symmetry, and can be verified from (\ref{currentsN4}) using (\ref{eomN4}).

\subsection{The \texorpdfstring{$\N=4$}{N=4} Weyl multiplet and its linear variations}\label{SSec:The N=4 Weyl multiplet and its linear variations}
 We are now able to apply the supersymmetry variations of the SYM multiplet to the currents that were found. We will find that to close the superalgebra on the Noether currents new (matter) currents have to be introduced. The full multiplet of currents will then give rise to the Weyl multiplet and its linear supersymmetry variations, as will be explained in more detail below.

The variation of the stress energy tensor is given by\footnote{The special supersymmetry with parameter $\eta $ in (\ref{Eq:SUSY variations of N=4 SYM}) and  (\ref{epsilonineta}) is neglected in the linear transformations considered here. It will be restored later using the superconformal algebra.}
\begin{align}
\susyvar \curstr{\mu\nu}{} = -\frac{1}{2} \susyparrc{}{k} \gammas{\rho (\mu}{} \der{}{\rho} \cursup{\nu)k}{} + \hc.
\end{align}
This is something that could be expected since it says that the graviton and gravitino are related through supersymmetry variations. However, when the supersymmetry current is varied one finds something new, namely:
\begin{equation}
\begin{aligned}
\susyvar \cursup{\mu i}{} =& -\frac{1}{2} \gammas{}{\nu} \curstr{\mu \nu}{} \susyparr{i}{} - (\di \curv{\mu i }{\s \;j} + \frac{1}{3} \gammas{\mu \lambda}{} \der{}{\lambda} \gammas{}{\rho} \curv{\rho i}{\s \;j})\susyparr{j}{}\\
&- 2\left(\der{}{\lambda} \curt{\lambda \mu ij}{} + \frac{1}{3} \gammas{\mu \nu}{} \der{}{\nu} \gamma\cdot\curt{ ij}{}\right)\susyparr{}{j}\,,\qquad \gamma \cdot t_{ij}= \gammas{}{ab}  \curt{ab ij}{}
\end{aligned}
\end{equation}
Here $\curt{\mu\nu ij}{}$ is the first current not associated to a symmetry. Its explicit form in terms of the SYM multiplet is given by\footnote{The \textit{tilde symbol} $(\widetilde{t}_{ab})$ is used to denote the dualization of a field on the spacetime indices $[ab]$.}
\begin{align}
 t_{abij}=-2\varphi_{ij}F^-_{ab}+\ft14 \varepsilon _{ijk\ell}\bar \psi ^k\gamma_{ab}\psi ^\ell
\,,\qquad
\curt{ab ij}{}= -\tilde t_{ab ij}\,,\qquad t_{ab}{}^{ ij}=t^*_{abij}= \tilde t_{ab}{}^{ij}\,.
\label{currentt}
\end{align}
The rest of the currents in the supercurrent multiplet are found by consecutively applying these supersymmetry variations. This procedure will result in five more currents associated to matter fields. In terms of the SYM multiplet these new currents are given by
\begin{equation}
\begin{aligned}
   \curd{k\ell}{ij} =& \scalar{}{ij} \scalar{k\ell}{} - \frac{1}{6} \kron{k\ell}{ij} \scalar{}{mn} \scalar{mn}{} \,,
    \qquad \xi_k^{ij} = -\ft32\varphi ^{ij}\psi _k -\delta ^{[i}_k\varphi ^{j]\ell}\psi_\ell\,,\\
\cure{ij}{} =& \spinorc{i}{} \spinor{j}{}\,, \qquad \lambda_i = \frac{1}{2} \gammas{}{ab} \F{ab}{+} \spinor{i}{}\,, \qquad   c = \F{ab}{-}  \F{}{ab -}\,.
\end{aligned}
\end{equation}
They satisfy the following conditions
\begin{equation}
  e_{ij}=e_{ji}\,,\qquad \xi ^{ij}_k=-\xi ^{ji}_k\,,\qquad \xi ^{ij}_i=0\,,\qquad d^{ij}_{k\ell}= \ft14\varepsilon ^{ijmn}\varepsilon _{k\ell pq}d_{mn}^{pq}\,,\qquad d^{ij}_{ik}=0\,.
 \label{symmetrycurrents}
\end{equation}
The explicit supersymmetric variations of the currents are given by
\begin{align}
\susyvar d^{ij}_{k\ell} =& -\frac{4}{3}\left( \susyparrc{[k}{} \xi^{ij}_{\ell]} + \bar \epsilon ^{[i}\xi ^{j]}_{k\ell} +\bar \epsilon _n\delta ^{[i}_{[k}\xi _{\ell]}^{j]n}+\bar \epsilon^n\delta ^{[i}_{[k}\xi _{\ell]n}^{j]}\right)\,,\nonumber\\
\susyvar \xi^{ij}_k =&-\frac{3}{16}\gamma\cdot\curt{}{ ij}\susyparr{k}{}- \frac{1}{8}\delta _k^{[i}  \gamma\cdot\curt{}{ j]\ell} \susyparr{\ell}{} + \frac{3}{8} \levi{}{ijmn} \cure{mk}{}\susyparr{n}{}\nonumber\\
&- \frac{3}{4}\gamma^{\mu}  \curv{\mu k}{\s \;[i} \susyparr{}{j]}-\frac{1}{4}\delta _k^{[i}\gamma^{\mu} \curv{\mu \ell}{\s \;j]} \susyparr{}{\ell}
+ \frac{3}{4} \di \curd{k\ell}{ij} \susyparr{}{\ell},  \nonumber\\
\susyvar e_{ij} =&  \susyparrc{(i}{}\lambda_{j)} - \frac{2}{3} \levi{kmn(i}{} \susyparrc{}{k} \slashed{\partial}\xi^{mn}_{j)},  \nonumber\\
\susyvar \curt{\rho \sigma ij}{} =&  \susyparrc{[i}{} \gammas{[\rho }{} \cursup{\sigma] j]}{}  -  \frac{1}{4}\varepsilon _{ijk\ell} \susyparrc{}{k} \gammas{\rho \sigma}{}\lambda^{\ell}+ \frac{1}{3} \susyparrc{k}{} \di \gammas{\rho \sigma}{} \xi^{k}_{ij},  \nonumber\\
\susyvar \curv{\mu j}{\s \; i} =& - \frac{1}{2}\susyparrc{}{i} \cursup{\mu j}{}
+ \frac{1}{2}\bar \epsilon _{j} {\cal J}_{\mu }^i + \frac{1}{8}\delta^i_j (\bar \epsilon ^{k} {\cal J}_{\mu k}-\bar \epsilon _{k} {\cal J}_{\mu}^ k)+ \frac{2}{3} \bar \epsilon^{k} \gamma_{\mu \lambda}\partial^{\lambda} \xi^i_{kj} - \frac{2}{3} \bar \epsilon_{k} \gamma_{\mu \lambda}\partial^{\lambda} \xi_j^{ki}\,,\nonumber\\
\susyvar \lambda_i =&\frac{1}{2} c^* \epsilon_i
+\frac{1}{2} \slashed{\partial}e_{ik}\epsilon^k - \frac{1}{8}\varepsilon _{ijk\ell}\gammas{\mu\nu}{}(\di \curt{}{\mu\nu k\ell})  \susyparr{}{j},  \nonumber\\
\susyvar \cursup{\mu}{i}  =& -\frac{1}{2} \gammas{}{\nu} \curstr{\mu\nu}{} \susyparr{}{i}
+  \frac12 (\gammas{}{\rho} \gammas{\mu \lambda}{} -\frac13 \gammas{\mu \lambda}{} \gammas{}{\rho})\partial^{\lambda}\curv{\rho j}{\s \s i} \susyparr{}{j}  -\frac14 \der{}{\lambda}\left(\gamma\cdot \curt{}{ij} \gammas{\mu \lambda}{} + \frac{1}{3}  \gammas{\mu \lambda}{} \gamma\cdot \curt{}{ij}\right)\susyparr{j}{},  \nonumber\\
\susyvar c =&  \susyparrc{i}{} \slashed{\partial}\lambda^i\,,  \nonumber\\
\susyvar \curstr{\mu\nu}{} =&  -\frac{1}{2}\susyparrc{}{k} \gammas{\lambda (\mu}{}\der{}{\lambda} \cursup{\nu)k}{} + \hc\,.
\label{Eq:N=4 current variations}
\end{align}
One can check these by applying the known variation of the $\N=4$ SYM multiplet to the currents, which are themselves nonlinear combinations of the SYM fields. The lowest weight field in the multiplet is the $d$-current, which is in the $\mathbf{20}$ representation of the $\SU(4)$ R-symmetry group. Therefore, the multiplet is said to be in the $\mathbf{20}$ representation of $\SU(4)$.

With the extra \textit{matter-currents} the number of fermionic and bosonic components are equal, which is a necessary condition for linearized supersymmetry.

The gravity multiplet is found by coupling each of the currents to a corresponding field in the Lagrangian. This field will either be a gauge field or a matter field, depending on whether the current is related to a symmetry or not. In the explicit form of the coupling of currents to fields of the gravity multiplet we choose normalizations for the gauge/matter fields that are convenient for comparison with \cite{Bergshoeff:1980is}:
\begin{align}
\La_1 = & \curstr{\mu\nu}{} h^{\mu\nu} + 2 \gav{\mu j}{\s \;i} \curv{i}{\mu j}- \frac12 \gad{k\ell}{ij} \curd{ij}{k\ell} \nonumber\\
&+\left( C\,c+ \frac12 \gae{ij}{} \cure{}{ij}- \gat{ij}{\mu\nu} \curt{\mu\nu}{ij} -\gravitinoc{}{\mu i}\cursup{\mu i}{} - \bar \Lambda^i \lambda_i + \frac43 \gaxicc{k}{ij} \curxi{ij}{k} + \hc  \right)\,.
\label{L1N4}
\end{align}
In the current multiplet we saw that the $d$-current had the lowest weight. In the Weyl multiplet this means that the $D$-field will have the highest weight and is also in the $\mathbf{20}$ representation of $\SU(4)$.

The invariance requirement of the action leads, using the representation properties of these fields in Table \ref{Tab:N=4 Weyl multiplet and currents} and (\ref{symmetrycurrents}),
\begin{table}
\begin{center}
\begin{tikzpicture}
\clip node (m) [matrix,matrix of nodes,
fill=black!20,inner sep=0pt,
nodes in empty cells,
nodes={minimum height=1.17cm,minimum width=2.1cm,anchor=center,outer sep=0,font=\fontsize{10}{12}\sffamily},
row 1/.style={nodes={fill=black,text=white}},
column 1/.style={text width=2cm,align=center,every even row/.style={nodes={fill=black!2.5}}},
column 2/.style={text width=2cm,align=center,every even row/.style={nodes={fill=black!2.5}}},
column 3/.style={text width=4cm,align=center,every even row/.style={nodes={fill=black!2.5}},},
column 4/.style={text width=2cm,align=center,every even row/.style={nodes={fill=black!2.5}},},
column 5/.style={text width=2cm,align=center,every even row/.style={nodes={fill=black!2.5}},},
column 6/.style={text width=2cm,align=center,every even row/.style={nodes={fill=black!2.5}},},
row 1 column 1/.style={nodes={fill=black, text=white}},
prefix after command={[rounded corners=2mm]  (m.north east) rectangle (m.south west)}
] {
	Field & Gauge field & Properties & $\SU(4)$ repr. & Weyl weight & Chiral weight \\
	$c $	& $ C $ &  complex  & $ \mathbf{1} $ & $ 0 $ & $ 2 $ \\
	$ \lambda_{i} $	& $ \Lambda_{i} $ & $ P_L\Lambda_{i} = \Lambda_{i} $ & $ \mathbf{4} $ & $ \frac{1}{2} $ & $ \frac{3}{2} $ \\
	$ \cure{ij}{} $	&  $ \gae{ij}{} $ & symmetric \& complex & $ \mathbf{10} $ & $ 1$ & $ 1 $ \\
	$ \curt{ab}{ij} $	&   $ \gat{ab}{ij} $ & $ \gat{ab}{ij} = \gat{ab}{[ij]}, \quad \tilde{T}_{ab}^{ij} = - \gat{ab}{ij}  $ & $ \mathbf{6} $ & $ 1 $ & $ 1 $\\
	$ \xi_k^{ij} $	& $ \chi_k^{ij} $ & $\chi_k^{[ij]}= \chi_k^{ij}, \quad \chi _i^{ij}=0$, \newline$ P_L \chi_k^{ij}=\chi_k^{ij}$ & $ \mathbf{20} $ & $ \frac{3}{2} $ & $ \frac{1}{2} $\\
	$ \curd{kl}{ij} $	&  $\gad{kl}{ij} $ & $ \gad{[kl]}{[ij]} = \gad{kl}{ij}, \quad \gad{kj}{ij}=0 $ $ \gad{kl}{ij}=\frac{1}{4}\levi{}{ijmn}\levi{klpq}{}\gad{mn}{pq}$ & $ \mathbf{20'} $ & $ 2$ & $0 $\\
	$ \curstr{\mu\nu}{} $	&  $ \vb{\mu}{a} $ &  frame field  & $ \mathbf{1} $ & $ -1 $ & $ 0 $\\
	$ \cursup{\mu}{i} $	&  $  \gravitino{\mu}{i} $ & $ P_L \gravitino{\mu}{i} = \gravitino{\mu}{i} $ & $ \mathbf{4} $ & $ -\frac{1}{2} $ & $ \frac{1}{2} $\\
	$ \curv{\mu i }{\s \;j} $	&  $ \gav{\mu j}{\s \;i} $ & $\gav{\mu i}{\s \;i} = 0$ & $ \mathbf{15} $ & $ 0 $ & $ 0$ \\
};
\end{tikzpicture}
\caption{\it The multiplet of currents and their corresponding gauge/matter fields for $\N=4$ conformal supergravity. The third column shows some properties of the gauge fields derived from the properties of the currents, the fourth column gives the representation of the fields with respect to the $R$-symmetry group $\SU(4)$, and the fifth and sixth column respectively give the Weyl and chiral weights of the gauge/matter fields. Explanations for the chiral weights are given at the end of Sec. \ref{SSec:The N=4 Weyl multiplet and its linear variations}.}\label{Tab:N=4 Weyl multiplet and currents}
\end{center}
\end{table}
to the following linear variations of the fields
\begin{align}
  \susyvar h_{\mu \nu }  =& \frac12\bar \epsilon ^i\gamma _{(\mu }\psi _{\nu )i}+ \hc+ \partial _{(\mu }\xi _{\nu )}-\eta _{\mu \nu }\lambda_{\rm D} \,, \nonumber\\
\susyvar C =&  \frac{1}{2}\susyparrc{}{i} \Lambda_i ,\nonumber\\
\susyvar \gravitino{\mu}{i} =&-\frac12 \gamma ^{\rho \sigma  }\epsilon ^i\partial _\rho  h_{\sigma \mu }
- V_{\mu j}{}^i\epsilon ^j  - \frac{1}{4} \gamma\cdot\gat{}{ij} \gammas{\mu}{} \susyparr{j}{}+ \partial _\mu \epsilon ^i-\gamma _\mu \eta ^i\,,\nonumber\\
 \susyvar \Lambda _i=&\slashed{\partial }C\epsilon _i+\ft12E_{ij}\epsilon ^j+\ft14\varepsilon _{ijk\ell}\gamma\cdot T^{k\ell}\epsilon ^j\,,\nonumber\\
    \susyvar V_{\mu i}{}^j =& \bar \epsilon ^j\phi _{\mu i}- \bar \epsilon _i\phi _\mu ^j -\ft14\delta ^j_i \left(\bar \epsilon ^k\phi _{\mu k}- \bar \epsilon _k\phi _\mu ^k\right)
       -\ft1{2}\bar \epsilon ^k\gamma _\mu \chi _{ik}^j+\ft1{2}\bar \epsilon _k\gamma _\mu \chi ^{jk}_i + \partial _\mu \lambda _i{}^j\,, \nonumber\\
 \susyvar \gat{ab}{ij} =& 2 \susyparrc{}{[i} (\der{[a}{} \gravitino{b]}{j]} -  \gammas{[a}{} \phi_{b]}^{j]}) + \frac{1}{4} \susyparrc{}{k} \gammas{ab}{} \gaxi{k}{ij} + \frac{1}{8} \levi{}{ijk\ell} \susyparrc{k}{} \di \gammas{ab}{} \Lambda_{\ell} ,\nonumber\\
 \susyvar \gae{ij}{} =& \susyparrc{(i}{} \di \Lambda_{j)}  - \susyparrc{}{k} \gaxi{(i}{mn} \levi{j)kmn}{} , \nonumber\\
\susyvar \gaxi{k}{ij} =& -\frac{1}{4} \gammas{}{ab} \di  \gat{ab}{ij} \susyparr{k}{} - \frac{1}{4} \levi{}{ij\ell m} \di \gae{k\ell}{} \susyparr{m}{} + \frac{1}{2} \gad{k\ell}{ij} \susyparr{}{\ell} - \gammas{}{ab} \partial_{a}\gav{b k}{\s [i} \susyparr{}{j]} - \text{trace}, \nonumber\\
 \susyvar \gad{k\ell}{ij} =& -2 (\susyparrc{}{[i} \di \gaxi{k\ell}{j]} + \susyparrc{}{m} \di \kron{[k}{[i} \gaxi{\ell]m}{j]} ) + \hc \,,
\label{delfromL1N4}
\end{align}
where $\xi _\mu $, $\lambda _j{}^i$, $\epsilon ^i$, $\lambda _{\rm D}$ and $\eta ^i$ are undetermined due to the constraints (\ref{curvatureconstraints}). This leads to symmetries of the gauge fields. Furthermore we used in (\ref{delfromL1N4})
\begin{equation}
  \phi^{i}_{\mu} = \frac{1}{4} (\gammas{}{\rho \sigma } \gammas{\mu}{} - \frac{1}{3} \gammas{\mu}{} \gammas{}{\rho \sigma } ) \der{\rho }{} \gravitino{\sigma }{i}\,,
  \label{defphilinear}
\end{equation}
which transforms under the $\eta $ symmetry in the $\delta \psi_\mu ^i$ as $\delta \phi^{i}_{\mu} =\partial _\mu \eta ^i$.
\bigskip

In order to promote these transformations to the transformations of the gauge fields of the superconformal algebra, one interprets $h_{\mu \nu }$ as the first order perturbation of the frame field:
\begin{align}
\vb{\mu}{a} = \kron{\mu}{a} + h_{\mu\nu}\eta^{\nu a} + \cdots\,, \qquad \delta e_\mu ^a= - \lambda^{ab}\eta _{b\mu } \,,
\end{align}
which needs a compensating Lorentz transformation parametrized by $\lambda ^{ab}$ since $h_{\mu\nu}$ was defined to be symmetric, and $\vb{\mu}{a}$ is defined without a symmetry.

We introduce then the customary composite field
\begin{equation}
  \omega _\mu {}^{ab}(e)= 2 e^{\nu[a} \partial_{[\mu} e_{\nu]}{}^{b]} -
e^{\nu[a}e^{b]\sigma} e_{\mu c} \partial_\nu e_\sigma{}^c\,,
 \label{omegae}
\end{equation}
which in first order reduces to
\begin{equation}
  \omega _\mu {}^{ab}(e)=-2e^{a\rho }e^{b\sigma }\partial _{[\rho}h_{\sigma ]\mu }\,,\qquad \delta \omega _\mu {}^{ab}(e)= \partial _\mu \lambda ^{ab}+\ldots \,.
 \label{omegae1storder}
\end{equation}
This is thus the gauge field of the symmetry parametrized by $\lambda ^{ab}$, and the first term of the gravitino transformation in (\ref{delfromL1N4}) can then be written as
\begin{equation}
  \delta_Q (\epsilon )\psi _\mu^ i= \frac{1}{4}\omega _\mu {}^{ab}(e)\gamma _{ab}\epsilon ^i+\ldots \,.
 \label{delQpsiomega}
\end{equation}

When we consider then the gauge transformations in the first line of (\ref{delfromL1N4}), the $\xi _\mu $ transformation can, up to a $\lambda^{ab}$ transformation, be written as  $\delta e_\mu^a=\partial _\mu \xi^a$. The dilatation transformation is the first order in
\begin{equation}
  \delta _D(\lambda _{\rm D})e_\mu ^a = -\lambda _{\rm D}e_\mu ^a\,.
 \label{delDe}
\end{equation}
We explicitly introduce a gauge field $b_\mu $ with $\delta _D(\lambda _{\rm D})b_\mu =\partial _\mu\lambda _{\rm D} $ in the theory.
This gauge field however caries no degrees of freedom. The reason is that it is the only field that transforms under the special conformal symmetries. Specifically a special conformal transformation will only cause a shift in the $b_{\mu}$ field:
\begin{align}
\delta_K b_{\mu} = 2\lambda_{{\rm K} \mu}.
\label{delKbmu}
\end{align}
In order to have a covariant $\omega _\mu {}^{ab}$ under the local transformations  (\ref{delDe}), we have to use
\begin{equation}
  \omega _\mu {}^{ab}(e,b) = \omega _\mu {}^{ab}(e)+2e_\mu ^{[a}b^{b]}\,,
 \label{omegaeb}
\end{equation}
The covariantization of the gravitino transformations further requires that the $\partial_\mu \epsilon $ term is combined with a $b_\mu$ term. The linear transformation of the gravitino is therefore
\begin{align}
  \susyvar \gravitino{\mu}{i} =&D _\mu \epsilon ^i - \frac{1}{4} \gamma\cdot\gat{}{ij} \gammas{\mu}{} \susyparr{j}{}-\gamma _\mu \eta ^i\,, \nonumber\\
  D_\mu \epsilon ^i=&\left( \partial_\mu  +\ft12b_\mu +\ft14 \omega _\mu {}^{ab}(e,b)\gamma _{ab}\right)\epsilon ^i - V_{\mu j}{}^i\epsilon ^j\,.
 \label{delcovgravitino}
\end{align}
The two additions with $b_\mu$ terms sum up to an extra term $\ft12\gamma _\mu \slashed{b}\epsilon ^i$. Therefore, this is of the form\footnote{From the point of view of the final results this reflects the commutator \[\left[\delta_K (\lambda_{\rm K}), \susyvar\right] \gravitino{\mu}{i}=\delta_K (\lambda_{\rm K}) \susyvar\gravitino{\mu}{i}= 2\lambda_{{\rm K}\mu}\frac{\partial }{\partial b_\mu } \susyvar\gravitino{\mu}{i} = \delta _S(\eta^i = -\lambda_{{\rm K}\mu}\gamma ^\mu \epsilon ^i)\gravitino{\mu}{i}\,.\]}
of the undetermined $\eta $ term in  (\ref{delfromL1N4}) and thus does not invalidate the previous results. Hence we can include $b_\mu $ in the set of fields, including also the special conformal transformations.\footnote{The commutator of supersymmetry transformations on the new field $b_\mu $ determine at this level the contribution of special conformal transformations to the commutator of supersymmetries.}
\bigskip

At this time, we can identify the independent fields $e_\mu ^a$, $\psi_\mu^i$,  $V_{\mu i}{}^j$ and $b_\mu $ to gauge fields of the superconformal algebra  (\ref{Eq: Commutators of superconformal group})  (for ${\cal N}=4$ and without the $T$ symmetry) according to Table \ref{Tab:Superconformalgroup generators in four dimensions}.
\begin{table}
	\begin{center}
		\begin{tikzpicture}
		\clip node (m) [matrix,matrix of nodes,
		fill=black!20,inner sep=0pt,
		nodes in empty cells,
		nodes={minimum height=1cm,minimum width=2cm,anchor=center,outer sep=0,font=\fontsize{10}{12}\sffamily},
		row 1/.style={nodes={fill=black,text=white}},
		column 1/.style={text width=4cm,align=center,every even row/.style={nodes={fill=black!2.5}}},
		column 2/.style={text width=2cm,align=center,every even row/.style={nodes={fill=black!2.5}}},
		column 3/.style={text width=2cm,align=center,every even row/.style={nodes={fill=black!2.5}},},
		column 4/.style={text width=2cm,align=center,every even row/.style={nodes={fill=black!2.5}},},
		column 5/.style={text width=2cm,align=center,every even row/.style={nodes={fill=black!2.5}},},
		column 6/.style={text width=2cm,align=center,every even row/.style={nodes={fill=black!2.5}},},
		row 1 column 1/.style={nodes={fill=black, text=white}},
		prefix after command={[rounded corners=2mm]  (m.north east) rectangle (m.south west)}
		] {
			Transformation & Generator & Parameter & Gauge field & Independent components \\
			Translation	& $P_a$ &  $\zeta_a$  & $e_{\mu}^{a}$ & $ 5 $  \\
			Lorentz boosts & $ M_{ab} $ & $ \lambda_{ab} $ & $ \spin{\mu}{\s ab} $ & composite \\
			Dilatations	& $ D $ & $ \lambda_D $  & $ \dilaton{\mu}{} $ & $ 0 $ \\
			Special conformal &   $ K_a $ & $ \lambda_K^a  $ & $ \spcon{\mu}{\s a} $ & composite \\
			Chiral $\SU(\N)$ & $ U_i^{\s j} $ & $\lambda_i^{\s j}$ & $ \gav{\mu j}{\s \;i} $ & $ 3(\N^2-1) $\\
			Chiral $\U(1)$ &  $T $ & $ \lambda_T $  & $ \gaa{\mu}{} $ & $ 3 $ \\
			Supersymmetry &  $ Q^i $ & $\susyparr{i}{}$  & $ \gravitino{\mu}{i} $ & $ 12\N $ \\
			Special supersymmetry &  $  S^i $ & $ \eta_i $ & $ \spgravitino{\mu}{i} $ & composite\\
		};
		\end{tikzpicture}
		\caption{\it The symmetries of the superconformal group in four dimensions $\SU(2,2|\N)$, their generators, parameters and gauge fields. The final column denotes the number of independent components of the gauge fields. The curvature constraints  (\ref{Eq:Curvature constraints}) have already been imposed and the redundant gauge degrees of freedom have been subtracted in this component counting. The table also clearly shows that in the case of $\N>1$ one has to include matter fields in the full multiplet to ensure that the number of fermionic and bosonic components are equal.
}\label{Tab:Superconformalgroup generators in four dimensions}
	\end{center}
\end{table}
The composite fields (\ref{omegaeb}),  (\ref{defphilinear}) and
\begin{equation}
  f_\mu {}^a= -\frac12e^\nu _b\partial _{[\mu } \omega_{\nu ]}{}^{ab}(e,b)+\frac{1}{12}e_\mu ^ae^{\rho }_c e^\sigma_d\partial _{\rho  } \omega_{\sigma }{}^{cd}(e,b)\,,
 \label{fdeflinear}
\end{equation}
provide the (linearized) gauge fields for the other superconformal symmetries.
These composite fields can be understood as resulting from constraints on curvatures of the full superconformal algebra:
\begin{align}\label{Eq:Curvature constraints}
R_{\mu\nu}(P^a) =0\,, \qquad \gammas{}{\nu} \widehat{R}_{\mu\nu}(Q^i) = 0\,,\qquad \vb{b}{\nu}\widehat{R}_{\mu\nu}(M^{ab}) = 0\,.
\end{align}
So far, we used only the linearized form of these constraints, but further on we will use these constraints to complete (\ref{omegaeb}), (\ref{defphilinear}) and  (\ref{fdeflinear}) at the nonlinear level.
The hats on top of the curvatures imply that they are then dependent on the matter fields in the multiplet as well. The way that one determines these curvatures and the gauging of the Poincar\'e group is explained in more detail in \cite[chapter 11]{Freedman:2012zz}. A discussion on the curvature constraints is given in chapter 16 of the same book. Later on we will come back to the curvatures and give explicit expressions when we reduce to the $\N=3$ Weyl multiplet.
An important motivation for the constraints is that they reconcile general coordinate transformations with gauged translations.

For the independent fields, the linearized transformations are in (\ref{delfromL1N4}), with the completion  (\ref{delcovgravitino}) and $h_{\mu \nu }$ is replaced by $e_\mu ^a$ and $b_\mu $, which transform under supersymmetry as
\begin{align}
  \susyvar \vb{\mu}{a} =& \frac{1}{2} \susyparrc{}{i} \gammas{}{a} \gravitino{\mu i}{} + \hc\,, \nonumber\\
  \susyvar b_{\mu} =& \frac12 \susyparrc{}{i} \phi_{\mu i} + \hc.
\label{delsusyeb}
\end{align}
All the variations of the gauge fields are now consistent with what one would expect from a gauge theory, i.e.
\begin{align}
\delta(\epsilon) A_{\mu}^A = \der{\mu}{} \epsilon^A + \epsilon^C B_{\mu}^B f_{BC}^{\s \s A}\,,
\label{generaldelAmu}
\end{align}
for the algebra of the P$\SU(2,2\vert4)$ symmetry group. This also implies transformations under $S$-supersymmetry, dilatations, Lorentz and $\SU(4)$ transformations.
All the currents, together with their respective gauge/matter fields, chiral and Weyl weights, are listed in Table~\ref{Tab:N=4 Weyl multiplet and currents}.
This contains also chiral weights. To explain these we should now come back to the remark after (\ref{Eq: Commutators of superconformal group}), namely that the original SYM multiplet did not have assigned chiral weights. It was mentioned that the $T$ symmetry for ${\cal N}=4$ is not part of the simple group P$\SU(2,2|4)$. So far it has also not been included in the symmetries that have been gauged, and e.g. in the range of the index $A$ in  (\ref{generaldelAmu}). However, the currents in (\ref{currentsN4}),  (\ref{currentt}) and (\ref{symmetrycurrents}) only depend on the (anti-)self-dual $F^\pm_{\mu \nu }$ and not on $B_\mu$ explicitly. As was mentioned after \eqref{Eq: Commutators of superconformal group}, this allows one to assign chiral weights to the fields and supersymmetry parameters\footnote{A chiral weight $c$ of a field $\phi $ means that the field transforms as $\delta _T\phi =\rmi c \phi \lambda _T$, where $\lambda _T$ is the parameter of this transformation.\label{fn:c}}
\begin{equation}
  c(\varphi _{ij})=0\,,\qquad c(\psi_i)=-c(\psi ^i)=\frac{1}{2}\,,\qquad c(F^\pm_{\mu \nu })=\pm 1\,,\qquad c(\susyparr{}{i})= -c(\susyparr{i}{}) = \frac{1}{2}\,.
 \label{cSYM}
\end{equation}
We would like to remark that these weights are consistent with the transformations of the bonus symmetry discussed in \cite{Intriligator:1998ig}.\footnote{Note, however, that this is not a symmetry of the action (\ref{LagrN4SYM}).} The weight of the supersymmetry parameter was directly assigned using the commutator $[T,Q]$ in (\ref{Eq: Commutators of superconformal group}). These lead to the weights of all the currents and, requiring  (\ref{L1N4}) to be invariant, to the weights of the fields of the Weyl multiplet \cite{Bergshoeff:1980is} given in Table \ref{Tab:N=4 Weyl multiplet and currents}.

This bonus $\U(1)$ symmetry is not gauged so far, though we will come back to its gauging in Sec. \ref{SSec:The nonlinear variations of the N=4 Weyl multiplet}.

\subsection{The nonlinear variations of the \texorpdfstring{$\N=4$}{N=4} Weyl multiplet}\label{SSec:The nonlinear variations of the N=4 Weyl multiplet}

The next step in constructing the nonlinear supersymmetry variations consists in the covariantization of the linear ones.\footnote{Details of this procedure have been discussed in \cite[sec. 2]{Bergshoeff:1986mz}.}
 This comes down to a three step procedure:
\begin{enumerate}
	\item make the parameters local: $\susyparr{}{} \rightarrow \susyparr{}{}(x)$,
	\item change derivatives into covariant derivatives: $\der{\mu}{} \rightarrow \D{\mu}{} $,
	\item replace all derivatives of gauge fields with curvatures: $\der{[\mu}{} A_{\nu]} \rightarrow \frac{1}{2} R_{\mu\nu}(A)$.
\end{enumerate}
After the covariantization one has to make an ansatz concerning the nonlinear terms in the variations. These terms have to be consistent with the representations and the Weyl and chiral weights of the different fields. The consistency with $\SU(4)$ means that the indices of the fields need to be correctly contracted on the right-hand side of the variations such that they match the indices on the left hand side. The terms in the right-hand side will also have to be combined in such a way that the sums of the Weyl and chiral weights of the fields and derivatives equal those of the Weyl and chiral weight of the field in the left-hand side of the variation. Finally, the coefficients appearing in front of the nonlinear terms have to be determined by imposing soft algebra relations to be determined by comparing the action of the commutator on the different fields. We will specify the relations below.

The chiral bonus symmetry, which was so far only a rigid symmetry, can be extended to a local symmetry \cite{Bergshoeff:1980is}, using a rewriting of the scalar field $C$ of the Weyl multiplet as projective coordinate of
the coset space $\SU(1,1)/\U(1)$ as in \cite{Cremmer:1979up}.\footnote{One could expect that the field $C$ is related to the bonus symmetry since its current $c=F_{ab}^-F^{ab-}$ allowed for the bonus symmetry in $\N=4$ SYM in the first place.} This means that the complex field $C$ is replaced by a complex constrained doublet $\{\Phi _\alpha\}=\{\Phi_1,\Phi_2\}$, constrained to $|\Phi_1|^2-|\Phi _2|^2 = 1$. This new doublet has now the following transformation laws
\begin{equation}
  \delta \Phi_\alpha = U_\alpha {}^\beta \Phi _\beta + \rmi \lambda _T \Phi _\alpha \,,
 \label{delPhialpha}
\end{equation}
where $\lambda _T$ is a local parameter, and $U_\alpha {}^\beta $  with $(U_\alpha {}^\beta)^* = -\eta _{\alpha \gamma }U_\delta {}^\gamma \eta ^{\delta \beta }$  and $U_\alpha {}^\alpha =0$ are rigid parameters. Appropriate insertions of $\Phi_\alpha$ allowed the authors of \cite{Bergshoeff:1980is} to write the Weyl multiplet with a local $\lambda _T$ chiral symmetry, acting on all the fields $\phi $ as in footnote \ref{fn:c}, with the weights $c$ given in Table \ref{Tab:N=4 Weyl multiplet and currents}. The gauge connection is the composite field
\begin{align}\label{Eq: U(1) gauge field in N=4 variation}
\varUpsilon_{\mu} = -\frac12\rmi \Phi^{\alpha}\stackrel{\leftrightarrow}{\partial}_{\mu} \Phi_{\alpha} - \frac14\rmi \overline{\Lambda}{}^{i}\gammas{\mu}{} \Lambda_{i},\qquad \Phi ^1 =(\Phi _1 )^*\,,\qquad \Phi ^2 =-(\Phi _2 )^*\,.
\end{align}
The second term is arbitrary and is introduced to simplify further formulae where $\varUpsilon_{\mu}$ appear. This is the case for all covariant derivatives, which have now a term
\begin{equation}
  D_\mu \phi =\ldots - \rmi c\varUpsilon_{\mu}\phi \,.
 \label{DmuphiwithU1}
\end{equation}
The relation with the formulation in terms of $C$ is made by gauge fixing, imposing the gauge condition $\Phi _1= (\Phi _1)^*$. This also fixes $\lambda _T$ in terms of the rigid parameters $U_1{}^1$ and $U_1{}^2$
\begin{equation}
  \lambda _T = \rmi U_1{}^1 + \ft12\rmi U_1{}^2 C -\ft12\rmi (U_1{}^2 C)^*\,,
 \label{lambdaTfixed}
\end{equation}
where we have that
\begin{equation}
  C= \frac{\Phi _2}{\Phi _1}\,.
 \label{defC}
\end{equation}
On the other hand we can then also write that
\begin{equation}
  \Phi _1 = \frac{1}{\sqrt{1-|C|^2}}\,,\qquad \Phi _2 = \frac{C}{\sqrt{1-|C|^2}}\,.
 \label{Phi12C}
\end{equation}
The variation of the original scalar with respect to the new parameters is
\begin{equation}
  \delta C = U_2{}^1 - 2 U_1{}^1 C- U_1{}^2 C^2\,.
 \label{deltaC}
\end{equation}
In the previous subsection we considered the linearized theory, which is also linear in $C$. Hence it contains only the rigid $U_1^1$ symmetry, related to the remaining constant value of $\lambda _T$ according to (\ref{lambdaTfixed}) as $U_1{}^1=-\rmi\lambda _T$. Hence we have
\begin{equation}
  \delta C = 2\rmi \lambda _T C\,,
 \label{delClinearized}
\end{equation}
agreeing with the weight~2 assigned to $C$ in Table \ref{Tab:N=4 Weyl multiplet and currents}. The reason one introduces the extra local symmetry and scalar doublet is that it puts strong restrictions on the nonlinear supersymmetry transformations.

 The full nonlinear variations of the Weyl multiplet, including the $S$-supersymmetry variations were found in \cite{Bergshoeff:1980is} and are given by\footnote{See the translation of conventions in Appendix \ref{app:translate}.}
  \begin{align}\label{Eq: non-linear susys N=4}
 \susyvart \vb{\mu}{a} =& \frac{1}{2} \susyparrc{}{i} \gammas{}{a} \gravitino{\mu i}{} + \hc, \nonumber\\
 \susyvart \Phi_{\alpha} =& -\frac{1}{2} \susyparrc{}{i} \Lambda_i \levi{\alpha\beta}{} \Phi^{\beta}, \nonumber  \\
     \susyvart \gravitino{\mu}{i} =& \D{\mu}{} \susyparr{}{i} - \frac{1}{4} \gamma\cdot\gat{}{ij} \gammas{\mu}{} \susyparr{j}{} - \frac{1}{2} \levi{}{ijk\ell}\Lambda_{\ell} \susyparrc{j}{} \gravitino{\mu k}{}  -   \gammas{\mu}{} \susyparrconformal{}{i}\,,
  \nonumber\\
 \susyvart \Lambda_i =& \levi{}{\alpha \beta} \Phi_{\alpha} \DS \Phi_{\beta} \susyparr{i}{} + \frac{1}{2} E_{ij} \susyparr{}{j} + \frac{1}{4} \levi{ijk\ell}{} \gamma\cdot\gat{}{k\ell} \susyparr{}{j}, \nonumber\\
  \susyvart b_\mu =& \frac{1}{2}\left(\susyparrc{}{i} \phi_{\mu i}-\bar \eta ^i\psi _{\mu i}+\hc\right)\,,\nonumber\\
  \susyvart \gav{\mu j}{\s \;i} =&  \susyparrc{}{i} \phi_{\mu j} + \frac{1}{2} \susyparrc{}{k} \gammas{\mu}{} \gaxi{kj}{i}  - \frac{1}{4} \gae{}{ik} \levi{jkmn}{} \susyparrc{}{m} \gravitino{\mu}{n} - \frac{1}{12} \gae{}{ik} \susyparrc{j}{} \gammas{\mu}{} \Lambda_k    \nonumber\\
 &+ \frac{1}{8} \levi{}{ik\ell m} \susyparrc{k}{} \gammas{ab}{} \gat{\ell j}{ab} \gammas{\mu}{} \Lambda_m
 + \frac{1}{6} \levi{\alpha \beta}{} \susyparrc{}{i} \gammas{\mu}{} \Phi^{\alpha} \DS \Phi^{\beta} \Lambda_j   \nonumber\\
 & -\frac{1}{2} \susyparrc{}{[i} \gammas{}{a} \gravitino{\mu [k}{} \overline{\Lambda}_{\ell]}  \gammas{a}{} \Lambda^{k]}
 -\gravitinoc{\mu}{i} \susyparrconformal{j}{} - \hc - \text{trace},  \nonumber\\
  \susyvart \gae{ij}{} =& \susyparrc{(i}{} \DS \Lambda_{j)}  - \susyparrc{}{k} \gaxi{(i}{mn} \levi{j)kmn}{}  - \frac{1}{2} \overline{\Lambda}_i \Lambda_j \susyparrc{k}{} \Lambda^{k} + \overline{\Lambda}_k \Lambda_{(i} \susyparrc{j)}{} \Lambda^{k} + 2 \susyparrconformalc{(i}{} \Lambda_{j)},
  \nonumber\\
   \susyvart \gat{ab}{ij} =& \susyparrc{}{[i} \hat{R}_{ab}^{j]}(Q) + \frac{1}{4} \susyparrc{}{k} \gammas{ab}{} \gaxi{k}{ij} + \frac{1}{8} \levi{}{ijk\ell } \susyparrc{k}{} \DS \gammas{ab}{} \Lambda_{\ell} - \frac{1}{12} \gae{}{k[i}\susyparrc{}{j]} \gammas{ab}{} \Lambda_k   \nonumber\\
 &- \frac{1}{6} \levi{}{\alpha\beta} \susyparrc{}{[i} \gammas{ab}{} \Phi_{\alpha} \DS \Phi_{\beta} \Lambda^{j]} - \frac{1}{4} \levi{}{ijk\ell} \susyparrconformalc{k}{} \gammas{ab}{} \Lambda_{\ell}, \nonumber\\
   \susyvart \gaxi{k}{ij} =& -\frac{1}{4} \gammas{}{ab} \DS  \gat{ab}{ij} \susyparr{k}{} - \frac{1}{4} \levi{}{ij\ell m} \DS \gae{k\ell }{} \susyparr{m}{} + \frac{1}{2} \gad{k\ell }{ij} \susyparr{}{\ell}  -  \frac{1}{2} \gammas{}{ab} \hat{R}_{ab}(\gav{k}{\s [i}) \susyparr{}{j]}   \nonumber\\
 & + \frac{1}{4} \gae{k\ell }{} \gae{}{\ell [i} \susyparr{}{j]} - \frac{1}{12} \levi{k\ell mn}{} \gae{}{\ell [i} ( \gamma \cdot \gat{}{j]n} \susyparr{}{m} +\gamma \cdot \gat{}{mn} \susyparr{}{j]} )   \nonumber\\
 & +   \frac{1}{4} \levi{}{ij\ell m} \levi{}{\alpha \beta} \Phi_{\alpha} \DS \Phi_{\beta}  \gammas{ab}{} \gat{k\ell }{ab} \susyparr{m}{} + \frac{1}{4} \gammas{}{a} \susyparr{n}{} \left( \levi{}{ij\ell n} \gaxicc{\ell k}{m}  - \frac{1}{2} \levi{}{ij\ell m} \gaxicc{\ell k}{n} \right) \gammas{a}{} \Lambda_m   \nonumber\\
 &+ \frac{1}{4} \susyparr{}{[i} \left( \overline{\Lambda}^{j]}  \DS \Lambda_k  + \frac{1}{2} \overline{\Lambda}_k \DS \Lambda^{j]} \right) - \frac{1}{4} \gammas{}{ab} \susyparr{}{[i} \left( \overline{\Lambda}^{j]}  \gammas{a}{}\D{b}{} \Lambda_k  - \frac{1}{2} \overline{\Lambda}_k \gammas{a}{}\D{b}{} \Lambda^{j]} \right)  \nonumber\\
 &- \frac{1}{24} \levi{}{ij\ell m} \Lambda_m \left[5 \susyparrc{\ell}{} (\gae{kn}{} \Lambda^{n} + 2 \levi{\alpha\beta}{} \Phi^{\alpha} \DS \Phi^{\beta} \Lambda_k )
 - \susyparrc{k}{} (\gae{\ell n}{} \Lambda^{n} + 2 \levi{\alpha\beta}{} \Phi^{\alpha} \DS \Phi^{\beta} \Lambda_{\ell}  )\right] \nonumber\\
  &-\frac{1}{4} \gamma\cdot\gat{}{ij} \gammas{c}{} \susyparr{[k}{} \overline{\Lambda}^{\ell} \gammas{}{c} \Lambda_{\ell]}
 - \frac{1}{4} \gamma\cdot\gat{}{\ell[i} \gammas{c}{} \susyparr{[k}{} \overline{\Lambda}^{j]}\gammas{}{c}\Lambda _{\ell]} + \frac{1}{4} \susyparr{}{[i}\overline{\Lambda}^{j]} \Lambda^{m} \overline{\Lambda}_{k} \Lambda_m - \text{trace}\nonumber\\
 &+ \frac{1}{2} \gamma\cdot\gat{}{ij} \susyparrconformal{k}{} + \frac{1}{3} \kron{k}{[i} \gamma\cdot \gat{}{j]\ell}\susyparrconformal{\ell}{} -\frac{1}{2} \levi{}{ij\ell m} \gae{k \ell}{} \susyparrconformal{m}{}\nonumber\\
  &+ \frac{1}{4} \gammas{a}{}\susyparrconformal{}{[i}\overline{\Lambda}_k \gammas{}{a} \Lambda^{j]} + \frac{1}{12} \gammas{}{a} \kron{k}{[i}\susyparrconformal{}{j]} \overline{\Lambda}_{\ell} \gammas{a}{} \Lambda^{\ell} - \frac{1}{12} \gammas{}{a} \susyparrconformal{}{ \ell}\overline{\Lambda}_{\ell} \gammas{a}{}\kron{k}{[i}\Lambda^{j]}\,,  \nonumber\\
  \susyvart \gad{k \ell}{ij} =& -2\susyparrc{}{[i} \DS \gaxi{k \ell}{j]}  +\frac{1}{2} \levi{}{ijmn} \susyparrc{}{p} \gat{k \ell}{}\cdot  (2 \gat{np}{} \Lambda_m + \gat{mn}{} \Lambda_p )  \nonumber\\
  & + \frac{1}{4} \levi{k \ell mn}{}  \susyparrc{}{[i} \left(-4\gae{}{j]p}  \gaxi{p}{mn} + \gamma\cdot\gat{}{mn} \DSlr \Lambda^{j]} + \frac{2}{3} \gae{}{j]m} \gae{}{np} \Lambda_p  \right.\nonumber\\
  &\phantom{+ \frac{1}{4} \levi{k \ell mn}{}  \susyparrc{}{[i} }\left.+\frac{4}{3} \gae{}{j]n} \levi{}{\alpha\beta} \Phi_{\alpha} \DS \Phi_{\beta} \Lambda^{m}   +\gamma\cdot\gat{}{mn} \Lambda_p \overline{\Lambda}^{j]} \Lambda^p \right)\nonumber\\
& + \susyparrc{}{[i} \left( \gammas{a}{} \gaxi{k \ell}{m} \overline{\Lambda}^{j]}  \gammas{}{a} \Lambda_m  -  \levi{}{\alpha \beta} \Phi_{\alpha} \DS \Phi_{\beta} \gamma\cdot \gat{k \ell}{} \Lambda^{j]}  +\frac{1}{3} \Lambda_{[k} \gae{ \ell]m}{} \overline{\Lambda}^{j]} \Lambda^{m} \right) \nonumber\\
 &+ \frac{1}{12} \susyparrc{}{[i}\gammas{ab}{} \levi{\alpha \beta}{} \Phi^{\alpha} \DS \Phi^{\beta} \Lambda^{j]}  \overline{\Lambda}_k \gammas{}{ab} \Lambda_{\ell}    - \text{trace} + \hc,
 \end{align}
where
\begin{equation}
 D_\mu \epsilon ^i=\left( \partial_\mu  +\ft12b_\mu +\ft14 \omega _\mu {}^{ab}(e,b,\psi )\gamma _{ab}-\ft12\rmi \varUpsilon_{\mu}\right)\epsilon ^i - V_{\mu j}{}^i\epsilon ^j\,,
 \label{Dmuepsilon}
\end{equation}
with composite fields $\omega_\mu {}^{ab}(e,b,\psi ) $ and $\phi _\mu $ as solutions to the constraints (\ref{Eq:Curvature constraints}):
\begin{equation}
\begin{aligned}
 \phi^i _a =&-\frac12 \gamma ^b \widehat{R}'^i_{ab}(Q)+\frac1{12}\gamma _a\gamma ^{bc}\widehat{R}'^i_{bc}(Q)\,,\\
 \widehat{R}'_{\mu \nu }(Q^i)=& 2 \left( \partial _{[\mu} +\ft12 b_{[\mu}
 +\ft14 \gamma _{ab}\omega  _{[\mu}{}^{ab}-\ft12\rmi \Upsilon _{[\mu} \right)
   \psi _{\nu ]}^i -2V_{[\mu j } {}^i\psi_{\nu ]} ^j
   \\
    & -\ft1 {2} \gamma^{ab} T_{ab}^{ij}\gamma_{[\mu}\psi _{\nu ]j} + \frac12 \levi{}{ijkl}\bar \psi_{\mu j} \psi_{\nu k} \Lambda_{\ell}\,,\\
  \widehat{R}_{\mu \nu }(Q^i)=& \widehat{R}'_{\mu \nu }(Q^i)-2\gamma _{[\mu }\phi _{\nu ]}^i \,.
 \label{solnphia}
\end{aligned}
\end{equation}
The covariant derivatives on fields are covariant for all symmetries. E.g. on $\Phi _\alpha $ this is
\begin{equation}
  D_\mu \Phi _\alpha =  \left(\partial _\mu -\rmi \varUpsilon_{\mu}\right)\Phi _\alpha +\frac{1}{2} \bar \psi _\mu ^i \Lambda_i \levi{\alpha\beta}{} \Phi^{\beta},
 \label{DaPhi}
\end{equation}
but this simplifies in the expression
\begin{align}\label{Eq: U(1) gauge field in N=4}
\levi{}{\alpha \beta} \Phi_{\alpha} D_a \Phi_{\beta} = \levi{}{\alpha \beta} \Phi_{\alpha} \partial_a \Phi_{\beta} - \frac12 \gravitinoc{a}{i} \Lambda_{i}.
\end{align}
Note that this is the expression that reduces to $\partial _aC$ in the linearized theory.

The only elementary field that transforms under $K$ is $b_\mu $ as in (\ref{delKbmu}).

 The algebra of $\N=4$ conformal supergravity is a soft algebra, i.e. with parameters that are dependent on the fields in the multiplet. The algebra of $Q$ supersymmetries is
 \begin{align}
 \comm{\delta_Q(\susyparr{1}{})}{\delta_Q(\susyparr{2}{})} =& \delta_{\rm cgct}(\xi^{\mu}) + \delta_{M}(\lambda^{ab}_1) + \delta_Q(\susyparr{3}{i}) + \delta_S(\susyparrconformal{1}{i}) + \delta_{\SU(4)}(\lambda_{1j}^{\s \;i}) + \delta_{\U(1)}(\lambda_T) + \delta_{K}(\lambda_{1K}^a),\nonumber\\
 \xi^{\mu} =& \frac{1}{2} \susyparrc{2}{i} \gammas{}{\mu} \susyparr{1i}{} + \hc,  \nonumber\\
 \lambda^{ab}_1 = & -\susyparrc{1}{i} \susyparr{2}{j} \gat{ij}{ab} + \hc,
  \nonumber\\
 \susyparr{3}{i} =& \frac{1}{2} \levi{}{ijk\ell} \Lambda_j \susyparrc{1k}{} \susyparr{2\ell}{},  \nonumber\\
\susyparrconformal{1}{i} =& -\frac{1}{2} \susyparrc{1}{k} \susyparr{2}{\ell} \chi^i_{k \ell} - \frac{1}{8} (\susyparrc{2}{k} \gammas{a}{} \susyparr{1j}{} + \hc )(\gammas{}{a} \chi^{ij}_k + \frac{1}{2} \levi{}{ij \ell m} \gamma\cdot \gat{km}{} \gammas{}{a} \Lambda_\ell)  \nonumber\\
&-\frac{1}{48} (\susyparrc{2}{i} \gammas{a}{} \susyparr{1j}{} - \kron{j}{i} \susyparrc{2}{\ell}\gammas{a}{} \susyparr{1\ell }{} + \hc)\gammas{}{a}(\gae{}{jk} \Lambda_k - 2 \levi{}{\alpha \beta} \Phi_{\alpha} \DS \Phi_{\beta} \Lambda^j)  \nonumber\\
 &+\frac{1}{8} \levi{}{ijk \ell} \susyparrc{1k}{}\susyparr{2 \ell}{} \DS \Lambda_j+ \frac{1}{6}\susyparrc{1}{[i}\susyparr{2}{j]}(\gae{jk}{}\Lambda^k + 2 \levi{\alpha\beta}{} \Phi^{\alpha} \DS \Phi^{\beta}\Lambda_j),
  \nonumber\\
 \lambda_{1j}^{\s \;i} =& \frac{1}{4}\gae{}{ik} \levi{k \ell mj}{} \susyparrc{2}{\ell} \susyparr{1}{m} + \frac{1}{8}(\susyparrc{2}{k}\gammas{}{a}\susyparr{1j}{}+\susyparrc{2j}{}\gammas{}{a}\susyparr{1}{k} -
 \delta _j^k\bar \epsilon _{2\ell}\gamma _a\epsilon _1^\ell)\overline{\Lambda}^i \gammas{a}{} \Lambda_k  \nonumber\\
& -\frac{1}{8} \susyparrc{2}{i} \gammas{a}{} \susyparr{1j}{} \overline{\Lambda}^k\gammas{}{a}\Lambda_k - \hc - \text{trace}, \nonumber\\
\lambda_T =& \frac{1}{8}\rmi(\susyparrc{2}{i} \gammas{a}{} \susyparr{1 j}{} +\hc )(\overline{\Lambda}^j \gammas{}{a} \Lambda_i - \kron{i}{j} \overline{\Lambda}^k \gammas{}{a} \Lambda_k),  \nonumber\\
 \lambda_{1{\rm K}}^a =& \frac{1}{12} \susyparrc{2 i}{} \gammas{}{b} \susyparr{1}{j} \tilde {\hat{R}}_{ab}(\gav{ j}{\s i}) + \frac{1}{3} \susyparrc{2}{i}\susyparr{1}{j} \D{b}{} \gat{ij}{ab} + \frac{1}{32}\susyparrc{1i}{}\gamma\cdot \gat{jk}{}\gammas{}{a} \gamma\cdot  \gat{}{ij} \susyparr{1}{k}  \nonumber\\
& + \frac{1}{24}\rmi \susyparrc{2i}{} \gammas{b}{}\susyparr{1}{i} \levi{}{abcd}\left(\D{c}{}\Phi^{\alpha}\D{d}{} \Phi_{\alpha} - \frac{1}{4} (\overline{\Lambda}^j \gammas{c}{} \D{d}{} \Lambda_j - \hc )\right) + \hc,
\nonumber\\
\comm{\delta_S(\susyparrconformal{}{})}{\delta_Q(\susyparr{}{})} =& \delta_{D}(\lambda_D) + \delta_{M}(\lambda^{ab}_2) + \delta_{\SU(4)}(\lambda_{2j}^{\s \;i}) +  \delta_S(\susyparrconformal{2}{i}) + \delta_{K}(\lambda_{2{\rm K}}^a)\,,\nonumber\\
\lambda_D =& \frac{1}{2}\susyparrc{}{i}\susyparrconformal{i}{} + \hc,  \nonumber\\
 \lambda^{ab}_2 = & -\frac{1}{2}\susyparrconformalc{i}{}\gammas{}{ab} \susyparr{}{i} + \hc,
  \nonumber\\
\lambda_{2j}^{\s \;i} =& \susyparrc{}{i} \susyparrconformal{j}{} - \frac{1}{4} \kron{j}{i} \susyparrc{}{k} \susyparrconformal{k}{} - \hc,  \nonumber\\
\susyparrconformal{2}{i} =& \frac{1}{8} \levi{}{ijk\ell}\gammas{}{a} \Lambda_{\ell} \susyparrconformalc{k}{} \gammas{a}{} \susyparr{j}{} ,
  \nonumber\\
\lambda_{2{\rm K}}^a =&- \frac{1}{24} \susyparrconformalc{i}{} \gamma\cdot \gat{}{ij} \gammas{}{a} \susyparr{j}{} + \hc \,,\nonumber\\
\comm{\delta_S(\susyparrconformal{1}{})}{\delta_S(\susyparrconformal{2}{})} =&\delta_{K}(\lambda_{3{\rm K}}^a)\,,\nonumber\\
\lambda_{3{\rm K}}^a =& \frac{1}{2}\bar \eta _2^i \gamma ^a \eta _{1i}+\hc\,.
 \end{align}
This algebra is an extension of $\su(2,2|4)$ since it reduces to the latter upon putting the fields of the Weyl multiplet to zero.

\section{The \texorpdfstring{$\N=3$}{N=3} Weyl multiplet}\label{Sec:The N=3 Weyl multiplet}

\subsection{The truncation to \texorpdfstring{$\N=3$}{N=3} extended conformal supergravity}\label{Sec:Truncation to N=3}
In this section the construction of the Weyl multiplet for $\N=3$ supergravity in four dimensions will be discussed. The construction is based on an explicit breaking of the supersymmetric conformal group P$\SU(2,2|4)$ into $\SU(2,2|3)$. This breaking will happen in three steps. Firstly one supersymmetry parameter, $\susyparr{4}{}$, is set to zero and secondly the appropriate field components of the $\N=4$ Weyl multiplet will have to be found that keep an $\N=3$ symmetry after the truncation. Finally, the variations of the fields in the new representation are derived. Deriving the explicit supersymmetry variations will be the subject of the next section.

In our truncation we will find that the $\SU(4)$ $R$-symmetry group breaks to $\SU(3)\times \U(1)$. This means that the Latin indices used in the previous section will not run from $1$ to $4$ anymore, but from $1$ to $3$. For the linear transformations that we present first in this section, the $\U(1)$  factor provides the ${\cal N}=3$ $T$-symmetry in (\ref{Eq: Commutators of superconformal group}).
This will be more subtle for the nonlinear transformations since the non-linear $\U(1)$ R-symmetry is a combination of the bonus $\U(1)$ symmetry described in the previous section and the $\U(1)$ factor in the truncation $\SU(4)\rightarrow \SU(3) \times \U(1)$.
The way in which these two $\U(1)$ symmetries are merged was already described in \cite{Ferrara:1998zt}, and will be relevant when we will discuss the nonlinear theory in section \ref{SSec:The nonlinear supersymmetry transformations}.

In the truncation of the $R$-symmetry group we will have to choose a convention for the order of the Levi-Civita tensor. The following choice has been made:
\begin{align}
\levi{ijk}{} := \levi{ i j k 4}{}.
\end{align}
Considering the variations \eqref{Eq:N=4 current variations}, we can identify the set of currents containing $\Theta _{\mu \nu}$ that form a multiplet under the transformations $\epsilon _i$, $i=1,2,3$, putting $\susyparr{4}{}=0$. These are
\begin{align}
\text{Weyl}_c(\N=3) = \{\curd{n}{m},\, \xi_{ij},\, \xi^{i}, \, \cure{i}{}, \curt{i}{ab},\,\curv{\mu j }{\s \;i},\, \lambda_{(R,L)},\, \cursup{\mu i}{},\,\curstr{\mu\nu}{}\}\,,\qquad i=1,2,3,
\label{Eq:N=3 Weyl multiplet}
\end{align}
where we have defined
\begin{align}
\curd{n}{m} &:= \frac{1}{4}\levi{}{mk\ell}\levi{nij}{}\curd{k\ell}{ij}=d^{m4}_{n4}, \qquad \xi_{ij} := \frac12 \levi{k\ell(i}{} \xi^{k\ell}_{j)}\,, \qquad  \xi^{i} := \frac{1}{2} \xi^{ij}_{j}\,,\nonumber\\
\cure{i}{} &:= \cure{i4}{}, \qquad \curt{i}{ab} := \frac{1}{2}\varepsilon _{ijk}\curt{}{ab jk}=\tilde t_{i}^{ab}  \,, \qquad \curlR := \lambda_4\,, \qquad \curlL := \lambda^4\,.
\end{align}
The $\lambda$-current has become an $\SU(3)$ scalar and thus there is an ambiguity in the notation concerning the chirality of the spinor. To indicate the chirality of the spinor we use subscripts $L$ or $R$. For practical use, it is also useful to record the inverse of the transformation between the $\xi $ symbols:
\begin{equation}
  \xi ^{ij}_k = \varepsilon ^{ij\ell}\xi _{k\ell}-2\delta _k^{[i}\xi ^{j]}\,.
 \label{inversexitransf}
\end{equation}

This is a multiplet with $64+64$ components and is found to be in the $\mathbf{8}$ representation of the $\SU(3)$ R-symmetry group. This labeling stems from the fact that the lowest weight current, $d_n^m$, is in the $\mathbf{8}$ representation, having the traceless property, $\curd{n}{n} = 0$, inherited from the properties of the ${\cal N}=4$ current.
In \cite{Fradkin:1985am,vanNieuwenhuizen:1985dp} a similar suggestion has been made for the representations that should be present in the $\N=3$ Weyl multiplet. The next step is to truncate the supersymmetry transformations of \eqref{Eq:N=4 current variations}:
\begin{align}
\susyvar \curd{n}{m} =& -\frac{2}{3} \levi{}{ij m} \susyparrc{i}{}\curxi{j n}{} +2 \susyparrc{n}{} \curxi{}{m}-\frac{2}{3}\delta ^m_n \bar \epsilon _k\xi ^k + \hc , \nonumber\\
\susyvar \curxi{ij}{} =& -\frac{3}{16} \gamma\cdot  \curt{(i}{} \susyparr{j)}{} - \frac38 \cure{(i}{} \susyparr{j)}{}  -  \frac38 \levi{k\ell(i}{} \gammas{}{a} \curv{a j)}{\s \; k} \susyparr{}{\ell} + \frac{3}{4} \levi{k \ell (i}{} \di \curd{j)}{\ell} \susyparr{}{k} , \nonumber\\
\susyvar \curxi{}{i} =& -\frac{1}{32} \levi{}{ijk} \gamma\cdot \curt{k}{} \susyparr{j}{} - \frac3{16} \levi{}{ijk} \cure{j}{} \susyparr{k}{}  -  \frac1{16}  \gammas{}{a} \curv{a j}{\s \; i} \susyparr{}{j}+ \frac3{16}  \gammas{}{a} \curv{a j}{\s \; j} \susyparr{}{i} + \frac{3}{8}  \di \curd{j}{i} \susyparr{}{j} , \nonumber\\
\susyvar  \cure{i}{} =& \frac{1}{2} \susyparrc{i}{} \curlR -2 \levi{ijk}{} \susyparrc{}{j} \di \curxi{}{k} - \frac23 \susyparrc{}{j} \di \curxi{ij}{} ,  \nonumber\\
\susyvar \curt{i}{ab} =& \frac{1}{2} \levi{ijk}{} \susyparrc{}{j} \gammas{}{[a} \cursup{}{b]k} - \frac{1}{4} \susyparrc{i}{} \gammas{}{ab} \curlR - \frac{1}{3} \levi{ijk}{} \susyparrc{}{j} \di \gammas{}{ab} \curxi{}{k} + \frac13 \susyparrc{}{j} \di \gammas{}{ab} \curxi{ij}{}, \nonumber\\
\susyvar \curv{\mu j}{\s \;i} =& - \frac{1}{2}\susyparrc{}{i} \cursup{\mu j}{} + \frac{1}{8}\delta^i_j \susyparrc{}{k} \cursup{\mu k}{} - \frac{2}{3} \levi{jkl}{} \susyparrc{}{k} \gammas{\mu \lambda}{}\partial^{\lambda} \curxi{}{il} - \frac23 \susyparrc{}{i} \gammas{\mu \lambda}{} \partial^{\lambda} \curxi{j}{} + \frac23 \delta^{i}_{j} \susyparrc{}{k} \gammas{\mu\nu}{}\partial^{\nu} \curxi{k}{} - \hc, \nonumber \\
\susyvar \curlR =& \frac{1}{2} \di \cure{i}{} \susyparr{}{i} + \frac{1}{4} \gammas{ab}{}  (\di \curh{i}{ab}) \susyparr{}{i},\nonumber \\
\susyvar \cursup{\mu i}{}  =& -\frac{1}{2} \gammas{}{\nu} \curstr{\mu\nu}{} \susyparr{i}{} - \frac{1}{2}(\gammas{}{\rho} \gammas{\mu\nu}{} - \frac{1}{3} \gammas{\mu\nu}{} \gammas{}{\rho})\der{}{\nu} \curv{\rho i}{\s \;j} \susyparr{j}{}\nonumber\\
 & - \frac{1}{4} \levi{ijk}{} \der{}{\nu} \left(\gamma\cdot \curh{}{j}\gammas{\mu\nu}{} + \frac{1}{3} \gammas{\mu\nu}{} \gamma\cdot \curh{}{j}\right) \susyparr{}{k} , \nonumber\\
\susyvar \curstr{\mu\nu}{} =&  -\frac{1}{2}\susyparrc{}{k} \gammas{\lambda (\mu}{}\der{}{\lambda} \cursup{\nu)k}{} + \hc.
\end{align}
The coefficients of the different terms have been checked on consistency with the algebra. Namely, the following commutator, when acting on the currents, has been calculated in this procedure:
\begin{align}
\susycomm = \frac{1}{2} \susyparrc{2}{i} \gammas{}{\mu} \susyparr{1i}{} P_{\mu} + \hc.
\end{align}
As was already expected we find that the $R$-symmetry group partially breaks from $\SU(4)$ to $\SU(3) \times \U(1)$. This is supported by the fact that the associated current acquires a trace after truncation:
\begin{align}
\sum_{i = 1}^{4}\curv{\mu i}{\s \;i} = 0 \quad \Rightarrow \quad \sum_{i=1}^{3} \curv{\mu i}{\s \;i} = -\curv{\mu 4}{\s \;4} \neq 0.
\end{align}
Because we are interested in the irreducible representations of the symmetry group we split up the $R$-current in a traceless part and an $\SU(3)$ scalar, namely: $\curv{\mu j}{\s \;i}=
\mathfrak v_{\mu \,j}{}^i +\frac{1}{3}\rmi \delta _j^i \mathfrak a_\mu $:
\begin{align}\label{Eq: Splitting R-symmetry current}
\susyvar \mathfrak{v}_{\mu j}{}^{i} =& - \frac{1}{2}\susyparrc{}{i} \cursup{\mu j}{}+\frac{1}{6}\delta _j^i\susyparrc{}{k} \cursup{\mu k}{}  - \frac{2}{3} \levi{jk\ell}{} \susyparrc{}{k} \gammas{\mu \nu}{}\partial^{\nu} \curxi{}{i\ell}- \frac23 \susyparrc{}{i} \gammas{\mu \lambda}{} \partial^{\lambda} \curxi{j}{} + \frac29 \delta^{i}_{j} \susyparrc{}{k} \gammas{\mu\nu}{}\partial^{\nu} \curxi{k}{}  - \hc , \nonumber\\
\susyvar \mathfrak{a}_{\mu} =& \frac{1}{8}\rmi \susyparrc{}{i} \cursup{\mu i}{} - \frac{4\imag}{3} \susyparrc{}{i} \gammas{\mu \nu}{} \der{}{\nu} \curxi{i}{}  + \hc.
\end{align}
The current multiplet now consists of the following fields:
\begin{align}
\text{Weyl}_c(\N=3) = \{\curd{n}{m},\, \xi_{ij},\, \xi^{i}, \, \cure{i}{}, \curt{i}{ab},\,\mathfrak v_{\mu j }^{\phantom{\mu j}i},\,\mathfrak a_\mu,\, \lambda_{(R,L)},\, \cursup{\mu i}{},\,\curstr{\mu\nu}{}\}\,,\qquad i=1,2,3,
\label{Eq:N=3 Weyl current multiplet improved}
\end{align}

From the truncation of $\N=4$ conformal supergravity one finds that there are two (spin-$3/2$) multiplets in the $\mathbf{6}$ and $\bar{\mathbf{6}}$ representations of $\SU(3)$, additionally to the (spin-$2$) stress-energy tensor multiplet in the $\mathbf{8}$ representation of $\SU(3)$. The components of the $\N=4$ multiplet therefore decompose as $\mathbf{20} \rightarrow \mathbf{8} \oplus \mathbf{6} \oplus \mathbf{\bar{6}}$, under the truncation of the R-symmetry group ($\SU(4)\rightarrow\SU(3)\times\U(1)$).

The $\mathbf{6}$ and $\bar{\mathbf{6}}$ multiplets consist of $32+32$ components given by
\begin{align}
\Phi_{3/2} = \{ d_{ij}, \;  \xi^{i}_{j}, \; \hat{\xi}_{i}, \;  e^{ij}, \;e, \;  \hat{t}_{i}^{ab}, \;   v_{\mu}^{i},\; \lambda^i, \; \mathcal{J}^{\mu},\; c\} \label{Eq:Currents outside of Weyl}
\end{align}
and its conjugate multiplet\footnote{The conjugation has to be taken in the $\N=4$ representation.}, with
\begin{align}
&d_{ij} := \varepsilon_{ik\ell} d^{k\ell}_{j4}=d_{ji}\,,\qquad \xi^{i}_{j} := \xi^{i}_{j 4}, \qquad \hat{\xi}_{i} := \frac12 \varepsilon_{ijk} \xi^{jk}_4,\,\qquad  e := e_{44}\,, \nonumber \\
&\hat{t}_{i}^{ab} := t_{i4}^{ab}=-\tilde {\hat{t}}_{i}^{ab}\,,\qquad  v_{\mu}^{i} := v_{\mu 4}^{\s \s i}\,,\quad \mathcal{J}^{\mu} := \mathcal{J}^{\mu}_{4}\,.
\end{align}
The lowest weight component, $d_{ij}$, is in the $\mathbf{6}$ representation of $\SU(3)$. The supersymmetry variations of this multiplet are found in the same way as for the $\mathbf{8}$ representation, namely by a consistent truncation of the variations in the $\N=4$ multiplet:
\begin{align}
\susyvar \curd{kl}{} &= -\frac{4}{3} \susyparrc{(k}{} \xi_{l)} - \frac43 \levi{ij(k}{} \susyparrc{}{i}  \curxi{l)}{j}, \nonumber\\
\susyvar \xi_j^{i} &= -\frac{3}{16} \gamma\cdot\hat{t}_{j} \susyparr{}{i} + \frac{3}{8} \levi{imn}{} \cure{}{mj} \susyparr{}{n} + \frac{3}{8} \gammas{}{\mu}  \curv{\mu }{i}\susyparr{j}{} + \frac{3}{8} \levi{}{ikl} \di \curd{jl}{} \susyparr{k}{} - \text{trace} ,  \nonumber\\
\susyvar \hat{\xi}_{i} &= -\frac38  \cure{}{} \susyparr{i}{} - \frac38 \levi{ijk}{} \gammas{}{\mu} v_{\mu}^{j} \susyparr{}{k} + \frac38 \di \curd{ij}{} \susyparr{}{j} , \nonumber\\
\susyvar \cure{}{ij} &= \susyparrc{}{(i} \lambda^{j)} - \frac{4}{3} \levi{}{km(i} \susyparrc{k}{} \di \xi_{m}^{j)}, \nonumber\\
\susyvar \cure{}{} &= -\frac{4}{3}  \susyparrc{}{k} \di \hat{\xi}_{k}, \nonumber\\
\susyvar \hat{t}^{ab}_i &= \frac{1}{2} \susyparrc{i}{} \gammas{}{[a} \cursup{}{b] } - \frac{1}{4} \varepsilon_{ikl}\susyparrc{}{k}\gammas{}{ab} \lambda^{l} + \frac{1}{3} \susyparrc{j}{} \di \gammas{ab}{} \xi_{i}^{j}, \nonumber\\
\susyvar \curv{\mu}{i} &= -\frac12 \susyparrc{}{i} \cursup{\mu}{} + \frac{2}{3} \susyparrc{}{j} \gammas{\mu\nu}{} \der{}{\nu} \xi_{j}^{i} + \frac43\levi{}{ijk}\susyparrc{j}{} \gamma_{\mu \nu} \partial^{\nu} \hat{\xi}_k\,,\nonumber\\
\susyvar \lambda^i &= \frac12 c\,\susyparr{}{i} + \frac12 \di \cure{}{ij}\susyparr{j}{} - \frac14 \varepsilon^{ijk} \gamma_{ab} \di \hat{t}_{k}^{ab}, \nonumber\\
\susyvar \cursup{\mu}{} &= \frac{1}{2} (\gammas{}{\rho} \gammas{\mu \nu}{} - \frac{1}{3} \gammas{\mu\nu}{} \gammas{}{\rho}) \der{}{\nu} \curv{\rho}{i} \susyparr{i}{} + \frac{1}{4} ( \gammas{ab}{} \gammas{\mu\nu}{} + \frac{1}{3} \gammas{\mu\nu}{} \gammas{ab}{}) \der{}{\nu} \hat{t}^{ab}_{i}\susyparr{}{i}\nonumber\\
\susyvar c &= \susyparrc{i}{} \di \lambda^i\,.
\end{align}

Hence, we find that none of the currents in the $\N=3$ Weyl current multiplet are found in the right-hand side of the variations. This means that we can consistently put all of these currents to zero and conclude that the truncation of the $\mathbf{20}$ representation of $\SU(4)$ into $\mathbf{8}\oplus\mathbf{6} \oplus \bar{\mathbf{6}}$ of $\SU(3)$ is consistent and that the multiplets are disjoint:
\begin{equation}
\begin{aligned}
\susyvar \text{Weyl}_c(\N=3)&\subset \text{Weyl}_c(\N=3),\\
\susyvar \left( \Phi_{3/2},\,\bar{\Phi}_{3/2}\right) \cap &\text{Weyl}_c(\N=3) = 0.
\end{aligned}
\end{equation}

\subsection{Components of the Weyl multiplet(s)}\label{SSec:Components of the Weyl multiplet(s)}
Now that we know the content of the $\N=3$ Weyl current-multiplet, we can determine the gauge and matter fields in the Weyl multiplet.

The Weyl multiplet is simply found by coupling a field to each of the currents, exactly as was already done in the case of $\N=4$ conformal supergravity. The first order action then looks like
\begin{equation}\label{Eq: N3 action}
\begin{aligned}
S_1 (\N=3) = \int \rmd^4x \;&  \curstr{\mu \nu }{}h^{\mu\nu } + 2 \mathcal V_{\mu j}{}^i  \mathfrak v^\mu {}_ i{}^{j} -4 \gaa{\mu}{} \mathfrak a_{\mu}- \gad{n}{m} \curd{m}{n}\\
& + \left(\gae{}{i} \cure{i}{}- \gat{ab}{i} \curt{i}{ab}- \gravitinoc{\mu}{i} \cursup{i}{\mu}- \overline{\Lambda}_L \lambda_L
 +\frac43\gaxicc{}{ij} \curxi{ij}{} + 4\bar \zeta ^i \curxi{i}{} + \hc \right)\,,
\end{aligned}
\end{equation}
leading to the independent fields
\begin{align}
\text{Weyl}(\N=3) = \{h_{\mu \nu },\;\gravitino{\mu}{i},\; \galL, \;\mathcal V_{\mu j}^{\phantom{\mu j} \;i} ,\; \gaa{\mu}{},\; \gah{ab}{i} ,\;\gae{i}{}, \; \gaxi{ij}{}, \;\zeta _i,\;\;\gad{n}{m} \},
\label{WeylN3v1}
\end{align}
where $\mathcal V_{\mu j}^{\phantom{\mu j} \;i}$ is traceless. As in sec. \ref{SSec:The N=4 Weyl multiplet and its linear variations}, we can introduce then the Lorentz transformations, dilatations and special conformal transformations. This works exactly as explained there and leads to the replacement of $h_{\mu \nu }$ by $e_\mu ^a$, the introduction of the gauge field $b_\mu $ and the definition of composite gauge fields by the constraints (\ref{Eq:Curvature constraints}). This replaces the set  (\ref{WeylN3v1}) by
\begin{align}
\text{Weyl}(\N=3) = \{e_\mu ^a,\;\gravitino{\mu}{i},\; \galL, \;b_\mu ,\;\mathcal V_{\mu j}^{\phantom{\mu j} \;i} ,\; \gaa{\mu}{},\; \gah{ab}{i} ,\;\gae{i}{}, \; \gaxi{ij}{}, \;\zeta _i,\;\gad{n}{m} \}.
\label{WeylN3v2}
\end{align}
These fields are consistent with earlier suggestions made for the $\N=3$ Weyl multiplet in \cite{Fradkin:1985am,vanNieuwenhuizen:1985dp}, similar to the case of the current multiplet.
We will present the transformations in Sec. \ref{ss:N3lintransf}.

Although we have found the $\N=3$ Weyl multiplet through a reduction of the $\N=4$ current multiplet, which was then coupled to the gauge fields, one could also directly reduce the $\N=4$ Weyl multiplet.\footnote{We would like to mention that it is also possible to construct the $\N=3$ Weyl multiplet directly from the $\SU(2,2|3)$ algebra, which was done in \cite{Hegde:2018mxv} when version two of this paper was being finalized. This procedure however hides many of the subtleties and interesting connections between the different Weyl multiplets in extended conformal supergravity.} Both procedures are completely consistent. We have chosen the normalizations of the gauge fields that couple to the $\N=3$ current multiplet in such a manner that the Lagrangian given in \eqref{Eq: N3 action} resembles the one in \eqref{L1N4}. Considering the $\N=3$ Weyl multiplet as a reduced version of the $\N=4$ Weyl multiplet (setting the fields in (\ref{Eq:Currents outside of Weyl}) to zero), we can identify (\ref{L1N4}) with  (\ref{Eq: N3 action}) for
\begin{align}
&\Lambda_L = \Lambda_4, \qquad E_i = E_{i 4}, \quad T_i^{ab} = \levi{ijk}{} T^{jk ab},\qquad  \gaa{\mu}{}=-\frac{2}{3}\imag V_{\mu i}{}^i=\frac{2}{3}\imag V_{\mu 4}{}^4  
\,,\nonumber \\
&\mathcal V_{\mu i}{}^j =V_{\mu i}{}^j-\frac13\delta _i^jV_{\mu k}{}^k\,,\qquad
\zeta ^i = 2 \chi^{ij}_j ,\quad \chi_{ij} = \levi{k\ell (i}{}\chi_{j)}^{k\ell}, \quad D^m_n = \levi{ijn}{}\levi{}{k\ell m}D_{k\ell}^{ij},
\label{N3fieldslinfromN4}
\end{align}
where all the $i,j,\ldots$ indices refer now to an $\SU{(3)}$ representation and thus run from one to three.
Two inverse relations that will be useful in the future are
\begin{equation}
  \chi ^{ij}_k= \ft12\varepsilon ^{ij\ell}\chi _{k\ell}-\ft12\delta _k^{[i}\zeta ^{j]}\,,\qquad V_{\mu i}{}^j= \mathcal V_{\mu i}{}^j+\frac12\rmi\delta _i{}^j{\cal A}_\mu \,.
 \label{chiVinverse}
\end{equation}
The last equation implies also how the $\SU(4)$ transformations contribute to the $\SU(3)$ and $T$ transformations. In fact, it implies for the parameters
\begin{equation}
  \lambda _i{}^j = \hat \lambda _i{}^j+\frac12\rmi\delta _i{}^j\lambda _T\,,\qquad \lambda _4{}^4 = -\frac32\rmi\lambda _T\,.
 \label{SU4to3}
\end{equation}
The parameter $\lambda _i{}^j$ has a trace and corresponds to the gauge field $V_{\mu i}{}^j$, $\lambda _T$ is the tracepart and corresponds to the gauge field $\mathcal A_\mu$, finally the symbol $\hat \lambda _i{}^j$ is the traceless part of $\lambda _i{}^j$ and corresponds to the gauge field  $ \mathcal V_{\mu i}{}^j$. Because the notation is rather heavy we will omit the hat in the future and from here on assume that the R-symmetry parameter $\lambda_i {}^j$ is once again traceless.

We can then calculate the weights under this $\lambda _T$ transformation for all the fields in the ${\cal N}=3$ multiplet (\ref{WeylN3v2}), using their definition from ${\cal N}=4$ fields in (\ref{N3fieldslinfromN4}). It turns out that this weight is equal to the chiral weight of the extra generator in ${\cal N}=4$ mentioned in the last column of Table \ref{Tab:N=4 Weyl multiplet and currents}. Therefore we can identify these symmetries and use $\lambda _T$ for both. The corresponding gauge field will then be the sum of ${\cal A}_\mu $ in (\ref{N3fieldslinfromN4}) and $\Upsilon _\mu $ reduced from (\ref{Eq: U(1) gauge field in N=4 variation}). However, the latter has no linear part after the truncation to ${\cal N}=3$ and will only play a role in the non-linear theory. We will therefore come back to this in Sec. \ref{SSec:The nonlinear supersymmetry transformations}, see (\ref{Amunonlin}).

The summary of the currents and the fields for ${\cal N}=3$ is given in Table \ref{Tab:N=3 Weyl multiplet and currents}, which also mentions these chiral weights.

\begin{table}
	\begin{center}
		\begin{tikzpicture}
		\clip node (m) [matrix,matrix of nodes,
		fill=black!20,inner sep=0pt,
		nodes in empty cells,
		nodes={minimum height=1.17cm,minimum width=2.1cm,anchor=center,outer sep=0,font=\fontsize{10}{12}\sffamily},
		row 1/.style={nodes={fill=black,text=white}},
		column 1/.style={text width=2cm,align=center,every even row/.style={nodes={fill=black!2.5}}},
		column 2/.style={text width=2cm,align=center,every even row/.style={nodes={fill=black!2.5}}},
		column 3/.style={text width=3.8cm,align=center,every even row/.style={nodes={fill=black!2.5}},},
		column 4/.style={text width=2cm,align=center,every even row/.style={nodes={fill=black!2.5}},},
		column 5/.style={text width=2cm,align=center,every even row/.style={nodes={fill=black!2.5}},},
		column 6/.style={text width=2cm,align=center,every even row/.style={nodes={fill=black!2.5}},},
		row 1 column 1/.style={nodes={fill=black, text=white}},
		prefix after command={[rounded corners=2mm]  (m.north east) rectangle (m.south west)}
		] {
			Current & Gauge field & Properties & $\SU(3)$ repr. & Weyl weight & Chiral weight \\
			$ \curlL $	& $ \galL $ & $ P_L\Lambda = \galL $,  $ P_R\Lambda = \galR $ & $ \mathbf{1} $ & $ \frac{1}{2} $ & $ \frac{3}{2} $ \\
			$ \cure{i}{} $	&  $ \gae{i}{} $ & complex & $ \mathbf{3} $ & $ 1 $ & $ 1 $ \\
			$ \curh{ab}{i} $	&   $ \gah{ab}{i} $ & $  \tilde{T}_{ab}^i =  \gah{ab}{i} $ & $ \mathbf{3} $ & $ 1 $ & $ -1 $\\
			$ \curxi{ij}{} $	& $ \gaxi{ij}{} $ & $P_L \gaxi{ij}{} = \gaxi{ij}{} $ & $ \mathbf{6} $ & $ \frac{3}{2} $ & $ \frac{1}{2} $\\
                        $ \curxi{i}{} $        & $ \gaz{}{i} $ & $P_L \gaz{}{i} = \gaxi{}{i} $ & $ \mathbf{3} $ & $ \frac{3}{2} $ & $ \frac{1}{2} $\\
			$ \curd{n}{m} $	&  $\gad{n}{m} $ & $ \gad{n}{n}=0 $ & $ \mathbf{8} $ & $ 2$ & $0 $\\
			$ \curstr{\mu\nu}{} $	&  $ \vb{\mu}{a} $ &  frame field  & $ \mathbf{1} $ & $ -1 $ & $ 0 $\\
			$ \cursup{\mu}{i} $	&  $  \gravitino{\mu}{i} $ & $ P_L \gravitino{\mu}{i} = \gravitino{\mu}{i} $ & $ \mathbf{3} $ & $ -\frac{1}{2} $ & $ \frac{1}{2} $\\
			$ \mathfrak v_{\mu j }^{\phantom{\mu j} i} $	&  $ \mathcal V_{\mu j}^{\phantom{\mu j}i} $ & $\mathcal V_{\mu j}^{\phantom{\mu j}j} = 0$ & $ \mathbf{8} $ & $ 0 $ & $ 0$ \\
			$ \mathfrak a_{\mu}$	&  $ \gaa{\mu}{} $ & 
    & $ \mathbf{1} $ & $ 0 $ & $ 0$ \\
			$ - $	&  $ \dilaton{\mu}{} $ & dilaton & $ \mathbf{1} $ & $ 0 $ & $ 0$ \\
		};
		\end{tikzpicture}
		\caption{ \it The multiplet of currents and their corresponding gauge/matter fields for $\N=3$ conformal supergravity. The third column shows some properties of the gauge fields derived from the properties of the currents, the fourth column gives the $R$-symmetry representation of the currents (and fields) and the fifth and sixth column respectively give the Weyl and chiral weights of the gauge/matter fields. }\label{Tab:N=3 Weyl multiplet and currents}
	\end{center}
\end{table}

Now that we know the components of the $\N=3$ Weyl multiplet, we are able to summarize all the known Weyl multiplets in four dimensions. More information on these multiplets can be found in \cite{Bergshoeff:1980is,Howe:1981gz,Fradkin:1985am,vanNieuwenhuizen:1985dp}. The different known Weyl multiplets\footnote{For ${\cal N}=2$ there is also another version known, which has been called the 'dilaton Weyl multiplet' constructed first for $D=6$ and $D=5$  \cite{Bergshoeff:1986mz,Bergshoeff:2001hc}, and recently also for $D=4$ \cite{Butter:2017pbp}.} in four dimensions, varying from $\N=4$ to $\N=1$ are listed in Table \ref{tbl:WeylN1234}.
This table is an extension of Table \ref{tbl:USP2Nreps} given in the introduction. Here the representations are  further split into $\SU({\cal N})$ representations, which correspond to the fields of the Weyl multiplets as e.g. for ${\cal N}=4$ in Table \ref{Tab:N=4 Weyl multiplet and currents} and for ${\cal N}=3$ in Table \ref{Tab:N=3 Weyl multiplet and currents}.
Another systematic way of representing the states of the ${\cal N}\leq 4$ Weyl multiplets has been given in \cite{Ferrara:2018wqd}.

\begin{table}[htbp]
\newcommand{\repDnr}[2]{\begin{tabular}{@{}c@{}}(#1) \\ #2\end{tabular}}
\begin{center}
 $ \begin{array}{|cc|cc|cc|cc|cc|}\hline
     field & \# \mbox{states} &  \multicolumn{2}{c|}{{\cal N}=4}& \multicolumn{2}{c|}{{\cal N}=3}& \multicolumn{2}{c|}{{\cal N}=2}&  \multicolumn{2}{c|}{{\cal N}=1}\\
       &    &  \mbox{rep.}  & \# \mbox{states}&  \mbox{rep.} & \# \mbox{states}&  \mbox{rep.}& \# \mbox{states}&  \mbox{rep.} & \# \mbox{states}\\ \hline
    e_\mu {}^a & 5 & 1 & 5 & 1 & 5 & 1 & 5 & 1 & 5 \\ \hline
    V_\mu & 3 & \repDnr{101}{ 15} & 45 & \repDnr{11}{ 8} & 24 &\repDnr{2}{ 3} & 9 &   &   \\
    {\cal A}_\mu  & 3 &   &   & 1 & 3  & 1 & 3 & 1 & 3 \\
    T_{ab} & 3+3 & \repDnr{010}{ 6} & 36 & \repDnr{10}{ 3} & 18 & 1 & 6 &   &   \\ \hline
    D&1&  \repDnr{020}{ 20} &  20 & \repDnr{11}{ 8} & 8 &  1 & 1 &   &   \\
    C & 1+1 & 1 & 2 &   &   &     & &  &   \\
    E & 1+1 & \repDnr{200}{ 10} & 20 & \repDnr{10}{ 3} & 6 &     & &   &   \\ \hline\hline
\psi _\mu  & 4+4 & \repDnr{100}{ 4} & 32 & \repDnr{10}{ 3}& 24 &\repDnr{1}{ 2}& 16 & 1 & 8 \\\hline
    \chi & 2+2 & \repDnr{110}{ 20}& 80 & \repDnr{20)+(01}{ 6+3} & 36 &\repDnr{1}{ 2} & 8 &   &   \\
    \Lambda  & 2+2 & \repDnr{100}{ 4} & 16 & 1 & 4 & &     &   &   \\\hline\hline
      &   &   & 128+128 &   & 64+64 &  & 24+24 &   & 8+8 \\\hline\hline
  \end{array}$
\end{center} 
  \caption{\it Weyl multiplets, ordered according to massive spin. The names of the field contain only the spacetime indices, since the other indices depend on ${\cal N}$. The second column gives the off-shell number of components as representation of the little group $\SU(2)$. Adding the representation content (counting double the fields reducible under $\SU(2)$) agrees with the numbers in Table \ref{tbl:USP2Nreps}. For the nontrivial representations, also the Dynkin labels are given, which corresponds to the Young Tableaux (the $i$-th Dynkin label is the number of columns with $i$ vertical boxes). The fields that have two parts in the second column have also the conjugate representation in $\SU({\cal N})$.}\label{tbl:WeylN1234}
\end{table}

\section{Supersymmetry transformations of the \texorpdfstring{$\N=3$}{} Weyl multiplet}
\label{SSec:The gauge fields and their transformations}

To find the supersymmetry variations we follow the same procedure as was applied in the case of $\N=4$ conformal supergravity. This means that first the linear variations are determined by imposing invariance of the first order action, and inserting the superconformal symmetries, as explained at the end of the previous section. In a second step these variations will be covariantized. Thereupon an ansatz for the nonlinear terms in the variations will be made such that the representations, Weyl weights and chiral weights are consistent.  Finally, this will be checked for compatibility with the soft algebra, fixing the coefficients of the nonlinear terms.

\subsection{The linear supersymmetry transformations}
\label{ss:N3lintransf}
The procedure outlined at the end of the previous section leads to the following linear supersymmetry variations
	\begin{align}
	\susyvar \vb{\mu}{a} =& \frac{1}{2} \susyparrc{}{i} \gammas{}{a} \gravitino{\mu i}{} + \hc,
	\nonumber\\
	\susyvar \gravitino{\mu}{i} =& \D{\mu}{} \susyparr{}{i} - \frac18 \levi{}{ijk} \gamma\cdot  \gah{ k}{} \gamma _\mu \susyparr{j}{},
	\nonumber\\
\susyvar b_\mu =& \frac{1}{2}\bar  \epsilon ^i\phi _{\mu i}+ \hc
\nonumber\\
	\susyvar \mathcal V_{\mu j}^{\phantom{\mu j}i} =& \susyparrc{}{i} \phi_{\mu j}  - \frac{1}{8} \susyparrc{}{i} \gammas{\mu}{} \gaz{j}{} - \frac{1}{4} \levi{jk\ell}{} \susyparrc{}{k} \gammas{\mu}{} \gaxi{}{i\ell}  - \hc - \text{trace},
	\nonumber\\
	\susyvar \gaa{\mu}{} =& -\frac{1}{6} \imag \susyparrc{}{i} \spgravitino {\mu i}{} - \frac{1}{6} \imag \susyparr{}{i} \gammas{\mu}{} \gaz{i}{} + \hc , \nonumber\\
	\susyvar \galL =& \frac{1}{2} E_{i} \susyparr{}{i} - \frac{1}{4} \gammas{ab}{} \gat{i}{ab} \susyparr{}{i},
	\nonumber\\
	\susyvar \gae{i}{} =&  \frac{1}{2} \susyparrc{i}{} \di \galL + \frac34 \levi{ijk}{} \susyparrc{}{j} \gaz{}{k} + \frac12 \susyparrc{}{j} \gaxi{ij}{} , \nonumber\\
	\susyvar \gat{ab}{i} =& \frac{1}{4} \susyparrc{}{i} \di \gammas{ab}{} \galR + \levi{}{ijk} \susyparrc{j}{} \hat{R}_{ab}(Q_k) + \frac14 \susyparrc{j}{} \gammas{ab}{} \gaxi{}{ij} - \frac18 \levi{}{ijk} \susyparrc{j}{} \gammas{ab}{} \gaz{k}{}, \nonumber\\
	\susyvar \gaxi{ij}{} =& -\frac14 \gammas{ab}{} \di \gah{(i}{ab} \susyparr{j)}{} + \frac{1}{2} \levi{k\ell (i}{} \gammas{}{ab} \hat{R}_{ab}(\mathcal V_{ j)}^{\phantom{j)}\ell})\susyparr{}{k} + \frac12 \di \gae{(i}{} \susyparr{j)}{} + \frac14 \levi{k\ell (i}{} \gad{j)}{\ell} \susyparr{}{k},\nonumber\\
	\susyvar \gaz{}{i} =&  \frac12 \levi{}{ijk} \di \gae{j}{} \susyparr{k}{} - \frac{1}{12} \levi{}{ijk} \gammas{ab}{} \di \gah{k}{ab} \susyparr{j}{} - \frac{1}{6} \gammas{}{ab} \hat{R}_{ab}(\mathcal V_{ k}^{\phantom{k} i}) \susyparr{}{k} +\frac{2}{3}\rmi \gammas{}{ab} \hat{R}_{ab}(\mathcal{A}) \susyparr{}{i}  +\frac{1}{4} \gad{k}{i} \susyparr{}{k},  \nonumber\\
	\susyvar \gad{n}{m} =& \frac32 \susyparrc{}{m} \di \gaz{n}{} - \levi{ijn}{}\susyparrc{}{i}\DS\gaxi{}{jm} - \text{trace} + \hc,
	\end{align}
with $D_\mu \epsilon _i$ as in (\ref{delcovgravitino}), which after the split in (\ref{chiVinverse}) is
\begin{equation}
  D_\mu \epsilon ^i= \left(\partial_\mu  +\ft12b_\mu +\ft14 \omega _\mu {}^{ab}(e,b)\gamma _{ab}-\ft12\rmi{\cal A}_\mu \right)\epsilon ^i - \mathcal V_{\mu j}^{\phantom{\mu j} i}\epsilon ^j\,.
 \label{DmuepsilonN3linear}
\end{equation}

These variations agree also with the reduction of (\ref{delfromL1N4}),  (\ref{delcovgravitino}) and (\ref{delsusyeb}), and are consistent with the ${\cal N}=3$ supersymmetry algebra. The highest order component of the multiplet is $D_n^m$, which is in the $\mathbf{8}$ representation of $\SU(3)$.

The next step is to derive the special supersymmetry transformations generated by $S^i$. This derivation is similar to the derivation of the $Q$-supersymmetry variations, the difference is that in this case the last line in (\ref{Eq: Commutators of superconformal group}) is
used to determine the coefficients in the variations. The non-trivial linear special supersymmetry variations are given by
\begin{equation}
\begin{aligned}
\susyvarspecial \gravitino{\mu}{i} =& - \gammas{\mu}{} \susyparrconformal{}{i}\,,\\
\susyvarspecial \dilaton{\mu}{} =& -\frac{1}{2}  \gravitinoc{\mu}{i} \susyparrconformal{i}{} + \hc\,, \\
\susyvarspecial \gav{\mu j}{\s\;i} =& -\gravitinoc{\mu}{i} \susyparrconformal{j}{} - \text{trace} - \hc , \\
\susyvarspecial \gaa{\mu}{} =& \,\frac{1}{6} \imag \gravitinoc{\mu}{i} \susyparrconformal{i}{} + \hc, \\
\susyvarspecial E_i =& \,\susyparrconformalc{i}{}\galL, \\
\susyvarspecial \gah{ab}{i} =& -\frac12 \susyparrconformalc{}{i} \gammas{ab}{} \galR, \\
\susyvarspecial \gaxi{ij}{} =&\, \frac12 \gamma\cdot  \gah{(i}{} \susyparrconformal{j)}{} + \gae{(i}{} \susyparrconformal{j)}{} ,\\
\susyvarspecial \gaz{i}{} =& -\frac{1}{6} \levi{ijk}{} \gamma\cdot  \gah{}{j} \susyparrconformal{}{k} + \levi{ijk}{} \gae{}{j} \susyparrconformal{}{k}\,.
\end{aligned}
\end{equation}

\subsection{The nonlinear supersymmetry transformations and the soft algebra}\label{SSec:The nonlinear supersymmetry transformations}
To find the nonlinear part of the supersymmetry variations one has to follow the same method as was described in the case of $\N=4$ conformal supergravity.
A careful analysis results in the following nonlinear variations
	\begin{align}
	\susyvart \vb{\mu}{a} =& \frac{1}{2} \susyparrc{}{i} \gammas{}{a} \gravitino{\mu i}{} + \hc,
	 \nonumber\\
	\susyvart \gravitino{\mu}{i} =& \D{\mu}{} \susyparr{}{i} - \frac18 \levi{}{ijk} \gamma\cdot  \gah{ k}{} \gamma _\mu \susyparr{j}{} - \frac{1}{2} \levi{}{ijk} \galL\susyparrc{j}{} \gravitino{\mu k}{}
- \gammas{\mu}{} \susyparrconformal{}{i},
	 \nonumber\\
	\susyvart \dilaton{\mu}{} =& \frac{1}{2} (\susyparrc{}{i} \spgravitino{\mu i}{} -\gravitinoc{\mu}{i} \susyparrconformal{i}{}) + \hc,  \nonumber\\
	\susyvart \mathcal V_{\mu j}^{\phantom{\mu j}i} =& \susyparrc{}{i} \phi_{\mu j}  - \frac{1}{8} \susyparrc{}{i} \gammas{\mu}{} \gaz{j}{} - \frac{1}{4} \levi{jk\ell}{} \susyparrc{}{k} \gammas{\mu}{} \gaxi{}{i\ell} -  \frac{1}{4} \levi{k\ell j}{} \gae{}{i} \susyparrc{}{k} \gravitino{\mu}{\ell} + \frac{1}{8} \susyparrc{}{i} \gammas{a}{} \gravitino{\mu j}{} \galLc \gammas{}{a} \galR \nonumber\\
	& +\frac{1}{12} \susyparrc{}{i}\left( E_j+\frac{3}{4}  \gamma\cdot\gat{j}{}\right)\gammas{\mu}{} \galR  - \gravitinoc{\mu}{i} \susyparrconformal{j}{}   - \hc - \text{trace},
	 \nonumber\\
	\susyvart \gaa{\mu}{} =& -\frac{1}{6} \imag \susyparrc{}{i} \spgravitino {\mu i}{} - \frac{1}{6} \imag \susyparrc{}{i} \gammas{\mu}{} \gaz{i}{} + \frac{1}{6} \imag \levi{ijk}{} \gae{}{i} \susyparrc{}{j} \gravitino{\mu}{k}+ \frac{1}{6}\rmi \susyparrc{}{i} \gammas{a}{} \gravitino{\mu i}{} \galLc  \gammas{}{a} \galR\nonumber\\
	&
 + \frac{1}{9} \rmi \susyparrc{}{i}\left(\gae{i}{}+\frac{3}{4}\gamma\cdot \gat{j}{}\right)\gammas{\mu}{} \galR +  \frac{1}{6} \imag \gravitinoc{\mu}{i} \susyparrconformal{i}{} + \hc , \nonumber\\
	\susyvart \galL =& \frac{1}{2} E_{i} \susyparr{}{i} - \frac{1}{4} \gamma\cdot  \gat{i}{} \susyparr{}{i},
	 \nonumber\\
	\susyvart \gae{i}{} =&  \frac{1}{2} \susyparrc{i}{} \DS \galL + \frac34 \levi{ijk}{} \susyparrc{}{j} \gaz{}{k} + \frac12 \susyparrc{}{j} \gaxi{ij}{} + \frac{1}{2} \galLc \galL \susyparrc{i}{} \galR + \susyparrconformalc{i}{}\galL,  \nonumber\\
	\susyvart \gat{ab}{i} =& \frac{1}{4} \susyparrc{}{i} \DS \gammas{ab}{} \galR + \levi{}{ijk} \susyparrc{j}{} \hat{R}_{ab}(Q_k) + \frac14 \susyparrc{j}{} \gammas{ab}{} \gaxi{}{ij} -  \frac18 \levi{}{ijk} \susyparrc{j}{} \gammas{ab}{} \left(\gaz{k}{}-\frac{2}{3} \gae{k}{} \galR\right) \nonumber\\
	&- \frac12 \susyparrconformalc{}{i} \gammas{a b}{} \galR,  \nonumber\\
	\susyvart \gaxi{ij}{} =& -\frac14 \gammas{ab}{} \DS \gah{(i}{ab} \susyparr{j)}{} + \frac{1}{2} \levi{k\ell (i}{} \gamma\cdot  \hat{R}(\mathcal V_{ j)}^{\phantom{j)} \ell})\susyparr{}{k} + \frac12 \DS \gae{(i}{} \susyparr{j)}{} + \frac14 \levi{k\ell (i}{} \gad{j)}{\ell} \susyparr{}{k}  \nonumber\\
	& + \frac14   \levi{k\ell(i}{} \gae{}{k} \left(\gae{j)}{}-\frac{1}{2}\gamma\cdot  \gah{j)}{}\right) \susyparr{}{\ell} -\frac14 \gammas{}{a}\susyparr{(i}{}\bar{\zeta}_{j)} \gammas{a}{} \galL\nonumber\\
	&+\left(\frac16\gae{(i}{}-\frac18 \gamma\cdot \gat{(i}{} \right)\gammas{}{a} \susyparr{j)}{}\galRc \gammas{a}{} \galL + \frac{1}{2} \gamma\cdot \gah{(i}{} \susyparrconformal{j)}{} + \gae{(i}{} \susyparrconformal{j)}{},  \nonumber\\
	\susyvart \gaz{}{i} =&  \frac12 \levi{}{ijk} \DS \gae{j}{} \susyparr{k}{} - \frac{1}{12} \levi{}{ijk} \gammas{ab}{} \DS \gah{k}{ab} \susyparr{j}{}- \frac{1}{6} \gamma\cdot  \hat{R}(\mathcal V_{ k}^{\phantom{k} i}) \susyparr{}{k}  +\frac23 \rmi \gamma \cdot \hat{R}(\mathcal{A})\susyparr{}{i} +\frac{1}{4} \gad{k}{i} \susyparr{}{k} \nonumber\\
  & - \frac{1}{24} \gae{}{i} \gamma\cdot  \gah{j}{} \susyparr{}{j} + \frac{1}{8} \gae{}{j} \gamma\cdot  \gah{j}{} \susyparr{}{i}   + \frac{1}{6} \gae{j}{} \gae{}{(i} \susyparr{}{j)}
  \nonumber\\
  & + \frac18 \levi{}{ijk } \gammas{}{a} \susyparr{k}{} \bar{\zeta}_{j} \gammas{a}{} \galL + \frac{1}{12} \gammas{a}{} \susyparr{j}{} \gaxicc{}{ij} \gammas{a}{} \galL \nonumber\\
	& -\frac16 \susyparr{}{i} (\galRc \slashed{D} \galL  + \frac{1}{2} \galLc \slashed{D} \galR)+\frac12 \gammas{}{ab} \susyparr{}{i}(\galRc \gammas{a}{} \D{b}{} \galL-\frac{5}{6}\galLc \gammas{a}{} \D{b}{} \galR) \nonumber\\
	& - \frac1{12} \levi{}{ijk} \gamma _a\susyparr{j}{} \galRc \gamma ^a \galL \gae{k}{} - \frac16\susyparr{}{i} \galRc \galR \galLc \galL  \nonumber\\
	&+ \levi{}{ijk} \left( \gae{j}{} - \frac{1}{6}\gamma\cdot \gah{j}{}\right) \susyparrconformal{k}{} -\frac{1}{6} \gammas{}{a} \susyparrconformal{}{i}\galLc \gammas{a}{} \galR,  \nonumber\\
	\susyvart \gad{j}{i} =& \frac32 \susyparrc{}{i} \DS \gaz{j}{} -  \levi{jk\ell}{}\susyparrc{}{k}\DS\gaxi{}{i\ell} +  \levi{k\ell j}{} \susyparrc{}{k} \gaz{}{(i} \gae{}{\ell)} - \gae{}{k} \susyparrc{}{i} \gaxi{jk}{} \nonumber\\
	& + \frac{1}{4} \susyparrc{}{i} \gamma\cdot  \gah{j}{} \DSlr \galR  - \frac{1}{6} \levi{jk\ell}{} \gae{}{\ell} \gae{}{i} \susyparrc{}{k} \galL + \frac14 \susyparrc{}{i} \gamma\cdot \gat{j}{} \galL \galRc \galR   \nonumber\\
	&  - \frac12 \susyparrc{}{i} \gammas{a}{} \gaz{j}{} \galRc \gammas{}{a} \galL - \frac{1}{6} \gae{j}{} \susyparrc{}{i} \galL \galRc \galR \color{black}- \frac12 \levi{jk\ell}{} \susyparrc{}{k} \galL \gat{}{i}\cdot  \gat{}{\ell}\nonumber\\
	&- \text{trace} + \hc,
\label{nonlinearN3transf}
	\end{align}
where
\begin{equation}
  D_\mu \epsilon ^i = \left(\partial _\mu +\ft14\omega _\mu {}^{ab}(e,b,\psi )\gamma _{ab}+\ft12 b_\mu -\ft12\rmi{\cal A}_\mu \right)\epsilon ^i - \mathcal V_{\mu j}^{\phantom{\mu j} i}\susyparr{}{j}\,.
 \label{DmuepsN3}
\end{equation}
The ${\cal A}_\mu $ in here is the combination from the linear one in  (\ref{N3fieldslinfromN4}) and the reduction of $\Upsilon _\mu $ in  (\ref{Eq: U(1) gauge field in N=4 variation}). Note that the complex scalar $C$ describing the coset $\SU(1,1)/\U(1)$ vanishes in the truncation to the $\N=3$ Weyl multiplet. Therefore also $\Phi _\alpha $ disappears. As a result, the bonus $\U(1)$ symmetry described in Section \ref{SSec:The nonlinear variations of the N=4 Weyl multiplet} is not anymore an independent symmetry, instead, it recombines with the diagonal $\U(1)$ inside $\SU(4)$ to form the $\U(1)$ R-symmetry in $\SU(2,2|3)$.
As written after (\ref{SU4to3}), the weights of the fields for both the $\U(1)$ factor inside $\SU(4)$ and the extra generator $\lambda _T$ used in Section \ref{SSec:The nonlinear variations of the N=4 Weyl multiplet} are equal, and they can thus be combined in the new soft algebra. The gauge field associated to the $\U(1)$ R-symmetry in $\SU(2,2|3)$ will only inherit the fermionic term from the composite field $\Upsilon_\mu$ in \eqref{Eq: U(1) gauge field in N=4 variation}.
Explicitly one finds that the new $\U(1)$ gauge field in terms of the ${\cal N}=4$ fields is
\begin{equation}
  {\cal A}_\mu  = \frac23\rmi V_{\mu 4}{}^4 -\frac14\rmi\bar \Lambda^4\gamma _\mu \Lambda _4 = \frac23\rmi V_{\mu 4}{}^4 -\frac14\rmi\bar \Lambda_R\gamma _\mu \Lambda _L \,.
 \label{Amunonlin}
\end{equation}
Note that this relation also impacts the $\lambda_T$ parameter in the soft algebra. Furthermore, the curvature tensor of the $\mathcal A_\mu$ gauge field also gets corrected in the non-linear case:
\begin{align}
\hat{R}_{ab} (\mathcal A) =& \frac23\rmi  \hat{R}_{ab}(V_4^{\phantom{4}4}) + \frac12 \rmi \overline \Lambda_R \gamma_{[a}D_{b]} \Lambda_L- \frac12 \rmi \overline \Lambda_L \gamma_{[a}D_{b]} \Lambda_R\,.
\label{RAintermsofN4}
\end{align}
This correction for the curvature has already been applied in \eqref{nonlinearN3transf}. Explicit expressions of the curvature tensors will be given shortly.

The nonlinear variations mentioned above are consistent with the soft algebra of the following form
\begin{align}
	\comm{\delta_Q(\susyparr{1}{})}{\delta_Q(\susyparr{2}{})} =& \delta_{\rm cgct}(\xi^{\mu}) + \delta_{M}(\lambda^{ab}_1) + \delta_Q(\susyparr{3}{i}) + \delta_S(\susyparrconformal{1}{i}) + \delta_{\SU(3)}(\lambda_{1j}^{\s \;i}) + \delta_{\U(1)}(\lambda_{1T}) + \delta _K(\lambda _{1{\rm K}}^a)\,,\nonumber\\
	\xi^{\mu} =& \frac{1}{2} \susyparrc{2}{i} \gammas{}{\mu} \susyparr{1i}{}+ \hc ,\nonumber\\
	\lambda^{ab}_1 =&  -\frac{1}{2}e_i \gah{}{ab i} -\frac{1}{2}e^i\gah{ i}{ab}\,, \qquad e_i\equiv  \varepsilon _{ijk}\bar \epsilon _1^j\epsilon _2^k\,,\nonumber\\
	\susyparr{3}{i} =& \frac{1}{2}\galL e^i, \nonumber\\
	\susyparrconformal{1}{i} =&\frac18\DS \galL e^i -\frac14 \gaxi{}{ij} e_j+ \left(\frac14 \gaz{j}{}-\frac{1}{6} E_j \galR\right)\susyparrc{1}{[i} \susyparr{2}{j]}
-  \frac{1}{16}\levi{}{ij\ell}\gammas{}{a}\gaxi{k\ell}{}
(\susyparrc{2}{k}\gammas{a}{}\susyparr{1j}{} + \hc)
\nonumber\\
	& -  \frac{1}{16}
\left(
\gammas{}{a}\gaz{}{[i}+ \gamma\cdot T^{[i}\gamma ^a \Lambda_L
-\frac{2}{3}\gae{}{[i} \gammas{}{a}\galL\right)(\susyparrc{2}{j]}\gammas{a}{}\susyparr{1j}{} + \hc)\,,
\nonumber\\
	\lambda_{1j}^{\s \;i} =& \frac14 e_j \gae{}{i}   +\frac{1}{8} \susyparrc{1}{i}\gammas{a}{}\susyparr{2j}{}  \galRc \gammas{}{a} \galL - \hc - \text{trace}, \nonumber\\
	\lambda_{1T} =& -\rmi \frac16 e_i \gae{}{i}   + \rmi \frac{1}{6} (\susyparrc{1}{j}\gammas{a}{}\susyparr{2j}{} + \hc) \galRc \gammas{}{a} \galL +\hc, \nonumber\\
 \lambda_{1{\rm K}}^a =& \frac{1}{12} \susyparrc{2 i}{} \gammas{}{b} \susyparr{1}{j} \tilde {\hat{R}}_{ab}(\gav{ j}{\s i}) +\frac{1}{24}\rmi \susyparrc{2 i}{} \gammas{}{b} \susyparr{1}{i} \tilde {\hat{R}}_{ab}({\cal A})\nonumber\\
&-\frac{1}{6}e^iD_bT^{ab}_i +\frac{1}{64}\bar \epsilon _{2i}\gamma \cdot T^{[i} \gamma ^a\gamma \cdot T_j\epsilon _1^{j]} + \hc,
\nonumber\\
	\comm{\delta_S(\susyparrconformal{}{})}{\delta_Q(\susyparr{}{})} =& \delta_{D}(\lambda_D) + \delta_{M}(\lambda^{ab}_2) + \delta_{\SU(3)}(\lambda_{2 j}^{\s \;i}) + \delta_{\U(1)}(\lambda_{2T})+  \delta_S(\susyparrconformal{2}{i}) + \delta _K(\lambda _{2{\rm K}}^a)\,,\nonumber\\
	\lambda_D =&  \frac{1}{2}\susyparrc{}{i}\susyparrconformal{i}{} + \hc,  \nonumber\\
	\lambda^{ab}_2 =&-\frac{1}{2}\susyparrconformalc{i}{}\gammas{}{ab} \susyparr{}{i} + \hc, \nonumber\\
	\lambda_{2j}^{\s \;i} =& \susyparrc{}{i} \susyparrconformal{j}{} - \hc - \text{trace},\nonumber\\
	\lambda_{2T} =&- \frac{1}{6} \imag \susyparrc{}{i} \susyparrconformal{i}{} + \hc  \nonumber\\
	\susyparrconformal{2}{i} =&- \frac{1}{8} \levi{}{ijk} \gammas{}{a} \galL\susyparrc{j}{} \gammas{a}{} \susyparrconformal{k}{}, \nonumber\\
\lambda_{2{\rm K}}^a =&- \frac{1}{48}\varepsilon ^{ijk} \susyparrconformalc{i}{} \gamma \cdot T_k \gammas{}{a} \susyparr{j}{} + \hc \,,\nonumber\\
\comm{\delta_S(\susyparrconformal{1}{})}{\delta_S(\susyparrconformal{2}{})} =&\delta_{K}(\lambda_{3{\rm K}}^a)\,,\nonumber\\
\lambda_{3{\rm K}}^a =& \frac{1}{2}\bar \eta _2^i \gamma ^a \eta _{1i}+\hc\,.
\label{softalgN3}
	\end{align}
Note that the parameter for the T-transformation has been corrected for the fact that the $\U(1)$ R-symmetry gets combined with the additional $\U(1)$ symmetry present in the $\N=4$ theory when reducing to the case of $\N=3$ supersymmetry.

It is also useful to have the transformations of constrained gauge fields. Following the method in Sec. 16.1.3 of \cite{Freedman:2012zz}, we find
\begin{align}
  \delta_{Q,S}(\epsilon ,\eta ) \omega _\mu {}^{ab}=  & \delta _{\rm gauge}\omega _\mu {}^{ab}+ \delta _{{\cal M}}\omega _\mu {}^{ab}\,, \nonumber\\
  \delta _{\rm gauge}\omega _\mu {}^{ab} =& \frac12 \bar \epsilon^i \gamma ^{ab}\phi _{\mu i}+\frac12 \bar \eta ^i \gamma ^{ab}\psi _{\mu i}+\frac12 \varepsilon _{ijk}\bar \epsilon ^i\psi _\mu ^jT^{ab\,k} +\hc\,,\nonumber\\
  \delta _{{\cal M}}\omega _\mu {}^{ab}=& \frac12\bar \epsilon^i\gamma _\mu \widehat{R}^{ab}(Q)_i+\hc\,,\nonumber\\
  \delta_{Q,S}(\epsilon ,\eta ) \phi  _\mu^i=  &\delta _{\rm gauge}\phi  _\mu^i+ \delta _{{\cal M}}\phi  _\mu^i\,, \nonumber\\
  \delta _{\rm gauge}\phi  _\mu^i=&{\cal D}_\mu \eta ^i-f_\mu ^a\gamma _a\epsilon ^i+\frac{1}{8}\varepsilon ^{ijk}\gamma ^a\Lambda _L\left(\bar \psi _{\mu j}\gamma _a\eta _k-\bar \epsilon_j \gamma _a\phi_{\mu k} \right)\nonumber\\
  &+ \frac18\DS \galL \varepsilon ^{ijk}\bar \psi  _{\mu\,j}\epsilon _k -\frac14 \gaxi{}{ij} \varepsilon _{jk\ell}\bar \psi  _{\mu}^k\epsilon ^\ell+ \left(\frac14 \gaz{j}{}-\frac{1}{6} E_j \galR\right)\bar \psi _\mu ^{[i} \susyparr{}{j]}\nonumber\\
	& -  \frac{1}{16}\left(
\gammas{}{a}\gaz{}{[i}\delta ^{j]}_k+ \gamma\cdot T^{[i}\delta ^{j]}_k\gamma ^a
-\frac{2}{3}\gae{}{[i}\delta ^{j]}_k \gammas{}{a}\galL+\levi{}{ij\ell}\gammas{}{a}\gaxi{k\ell}{}
\right)(\susyparrc{}{k}\gammas{a}{}\psi _{\mu \,j}{} + \hc)
\,,\nonumber\\
  \delta _{{\cal M}}\phi  _\mu^i=&
  \frac{1}{16}
   \left(\frac{1}{3}\gamma_{\mu} \gamma^{ab} - \gamma^{ab} \gamma_{\mu}\right)\left(\rmi\widehat{R}_{ab}\left(\mathcal A\right)\epsilon ^i +2 \widehat{R}_{ab}\left(\mathcal{V}_j^{\phantom{j}i}\right) \epsilon^j \right)\nonumber\\
   &-\frac1{16} \varepsilon ^{ijk} \gamma^{ab} \epsilon_j \overline \Lambda _L \gamma_\mu  \widehat R_{ab}\left(Q_k\right)
    - \frac{1}{16}\varepsilon^{ijk}\left( \slashed{D} \gamma_{ab}\gamma_\mu - \frac{1}{3} \gamma_\mu \gamma_{ab} \slashed{D}\right)T^{ab}_k \epsilon_j\nonumber\\
    &+\frac{1}{32}  \gamma\cdot T^{[i} \gamma_\mu \gamma\cdot T_j \epsilon^{j]}  - \frac{1}{24} \varepsilon^{ijk}\gamma_\mu \gamma\cdot T_k\eta_j\,.
\label{delconstrainedgauge}
\end{align}
The commutator relations also determine the covariant curvatures before the constraints are used. We find e.g.
\begin{equation}
\begin{aligned}
\hat{R}_{\mu\nu}(Q^i) =&2\left(\partial_{[\mu}+ \frac14\omega _{[\mu} {}^{ab}(e,b,\psi )\gamma _{ab}+\frac12 b_{[\mu} -\ft12\rmi{\cal A}_{[\mu} \right)\psi _{\nu ]}^i - 2\mathcal V_{[\mu j}{}^i\psi _{\nu ]}^{j} \\
    & + \frac14 \levi{}{ijk} \gamma\cdot  \gah{ j}{}\gamma _{[\mu} \psi _{\nu ]\,k} + \frac{1}{2} \levi{}{ijk} \galL\bar \psi _{\mu \,j} \gravitino{\nu\, k}{}
- 2\gammas{[\mu}{} \phi _{\nu ]}^{i}\,,\\
\hat{R}_{\mu\nu}(M^{ab}) =& 2\partial_{[\mu}\omega _{\nu ]}{}^{ab}+2\omega _{[\mu }{}^{ac}\omega _{\nu ]c}{}^b-\left( \bar \psi _{[\mu} ^i \gamma ^{ab}\phi _{\nu] i}+\frac12 \varepsilon _{ijk}\bar \psi _\mu  ^i\psi _\nu ^jT^{ab\,k} +\bar \psi _{[\mu }^i\gamma _{\nu]} \widehat{R}^{ab}(Q)_i+\hc\right)\,,\\
\hat{R}_{\mu\nu}(\mathcal V_{ j}^{\phantom{j} i}) =&2\partial _{[\mu }\mathcal V_{\nu ] j}{}^{i}+2 {\cal V}_{[\mu \,j}{}^k{\cal V}_{\nu ]k}{}^i \\
&+\left(-2\bar \psi _{[\mu }^{i} \phi_{\nu] j} + \frac{1}{4} \bar \psi _{[\mu }^{i}\gammas{\nu]}{} \gaz{j}{}+ \frac{1}{2} \levi{jk\ell}{} \bar \psi _{[\mu }^{k} \gammas{\nu]}{} \gaxi{}{i\ell} + \frac{1}{4} \levi{k\ell j}{} \gae{}{i} \bar \psi _\mu ^{k} \gravitino{\nu}{\ell} \right.\\
	& \left.+ \frac{1}{6} E^i\bar \psi _{[\mu\,j }\gammas{\nu]}{} \galL +\frac{1}{8}\bar \psi _{[\mu\, j}\gamma\cdot \gat{}{i} \gammas{\nu]}{} \galL - \frac{1}{8} \bar \psi _{[\mu}^{i} \gammas{a}{} \gravitino{\nu] j}{} \galLc \gammas{}{a} \galR    -\, \hc - \text{trace}\right)\,,\\
\hat{R}_{\mu\nu}(\gaa{}{})=&2\partial _{[\mu }{\cal A}_{\nu ]}+\left[\frac{1}{3}\rmi \bar \psi _{[\mu }^i\phi _{\nu ]i}+ \imag\frac{1}{6} \levi{i jk}{} \gae{}{i} \gravitinoc{\nu}{j}\gravitino{\mu}{k}
- \frac{1}{6}\rmi \bar \psi^i_\mu \gamma^a \psi_{\nu i}\bar \Lambda_L \gamma^a \Lambda_R
+ \imag \frac13\gravitinoc{[\mu}{i} \gammas{\nu]}{} \gaz{i}{}\right.  \\
&\left.+\imag\frac{2}{9}\gae{}{i} \gravitinoc{i[\mu }{}  \gammas{\nu]}{} \galL
+\imag\frac{1}{6} \gravitinoc{i[\mu}{} \gamma\cdot \gat{}{i} \gammas{\nu]}{} \galL + \hc \right],\\
\hat R_{\mu\nu}(S^i) =&2 \left(\partial_{[\mu}+ \frac14\omega _{[\mu} {}^{ab}(e,b,\psi )\gamma _{ab}+\frac12 b_{[\mu} +\ft12\rmi{\cal A}_{[\mu} \right)\phi _{\nu ]}^i - 2\mathcal V_{[\mu j}{}^i\phi _{\nu ]}^{j} \\
&-2 f_{[\mu}^a \gamma_a \psi_{\nu]}^i
-\frac14 \varepsilon^{ijk} \psi_{[\mu j}\bar \phi_{\nu] k} \Lambda_L\\
&+ \frac18\DS \galL \varepsilon ^{ijk}\bar \psi  _{\mu\,j}\psi_{\nu\, k}-\frac14 \gaxi{}{ij} \varepsilon _{jk\ell}\bar \psi  _{\mu}^k\psi_\nu^\ell+ \left(\frac14 \gaz{j}{}-\frac{1}{6} E_j \galR\right)\bar \psi _\mu ^{[i} \psi_{\nu}^{j]}\\
&+ \frac{1}{16}\left(
\gammas{}{a}\gaz{}{[i}\delta ^{j]}_k+ \gamma\cdot T^{[i}\delta ^{j]}_k\gamma ^a
-\frac{2}{3}\gae{}{[i}\delta ^{j]}_k \gammas{}{a}\galL+\levi{}{ij\ell}\gammas{}{a}\gaxi{k\ell}{}
\right)\left(\psi_{[\mu}^k\gammas{a}{}\psi _{\nu] \,j}+\hc
 \right)\\
& + \frac{1}{8}
 \left(\frac{1}{3}\gamma_{[\mu} \gamma^{ab} - \gamma^{ab} \gamma_{[\mu}\right)
 \left(\rmi\widehat{R}_{ab }\left(\mathcal A\right)\psi _{\nu ]} ^i +2 \widehat{R}_{ab }\left(\mathcal{V}_j^{\phantom{j}i}\right) \psi _{\nu ]}^j \right)
 \\
  &+\frac1{8} \varepsilon ^{ijk} \gamma^{ab} \psi _{[\mu j} \overline \Lambda _L \gamma_{\nu ]}  \widehat R_{ab}\left(Q_k\right)
    + \frac{1}{8}\varepsilon^{ijk}\left(\frac{1}{3} \gamma_{[\mu} \gamma_{ab} \slashed{D}-\slashed{D} \gamma_{ab}\gamma_{[\mu}\right)T^{ab}_k \psi _{\nu ]\,j} \\
    &+\frac{1}{16}  \gamma\cdot T^{[i} \gamma_{[\mu} \gamma\cdot T_j \psi _{\nu ]}^{j]}  - \frac{1}{12} \varepsilon^{ijk}\gamma_{[\mu} \gamma\cdot T_k\phi _{\nu ]j}\,.
\label{curvaturesfromtransfo}
\end{aligned}
\end{equation}
Note that the variations (\ref{nonlinearN3transf}) immediately determine the variations of the different curvatures associated to the gauge symmetries.
The explicit formula for deriving these variations is given in \cite[chapter 11]{Freedman:2012zz}.

\section{Conclusion}
\label{ss:conclusion}
	In this paper the gravity multiplet for $\N=3$ conformal supergravity in four dimensions, also known as the $\N=3$ Weyl multiplet, has been constructed. The construction of this multiplet was done using three steps. First the known \textit{current multiplet}  of the $\N=4$ conformal supergravity theory was truncated to $\N=3$ supersymmetry. These currents are essential because they uniquely determine the Weyl multiplet. This one to one correspondence follows from the first order action in which each current is coupled to a field in the multiplet:
	\begin{align}\label{Eq:First order action conclusion}
	S_1 \propto \int field \times current.
	\end{align}
The breaking of supersymmetry explicitly consisted of three consecutive steps:
	\begin{enumerate}
		\item put one supersymmetry parameter to zero $\susyparr{4}{} \rightarrow 0$,
		\item determine the supersymmetry transformations restricted to the remaining three supersymmetries,
		\item find a subset of currents that transform internally for the remaining three supersymmetries.
		\end{enumerate}
		It was found that the subset of currents consisted of $64+64 \subset 128+128$ fermionic and bosonic components. The $128+128$ components refer to the case of $\N=4$ conformal supergravity. Inside the found multiplet $40$ fermionic and $32$ bosonic components were assigned to matter fields. These fields are generally necessary in extended supergravity to ensure an equal amount of fermionic and bosonic components. 
		
		The truncation thus split up the $128+128$ components in the original $\N=4$ conformal supergravity in two subsets. One of these subsets contained the gravity multiplet\footnote{This includes the graviton, gravitino and gauge fields associated to the $R$-symmetry.} and the other multiplet could be interpreted as a matter multiplet. For the truncation to be consistent the two subsets had to be completely disjoint. The gravity multiplet transforms internally under supersymmetries, and the matter multiplet does not act as a source into the gravity multiplet when the supersymmetries are applied. We found that this is indeed the case:

		To further argue that the correct multiplet of currents was found, remark that previous literature suggested that the Weyl multiplet of $\N=3$ conformal supergravity indeed had to consist of $64+64$ fermionic and bosonic components \cite{Fradkin:1985am,vanNieuwenhuizen:1985dp}.

        The first order coupling in \eqref{Eq:First order action conclusion} can be viewed as a perturbation around flat space. By imposing invariance of this first order action, one then systematically finds the linearized supersymmetry transformations. Determining these linear supersymmetry variations is the second step in the three step procedure. To make sure that the found variations were correct, their consistency with the superconformal algebra was checked.

		The third step in the procedure was to determine the nonlinear variations. In this step the large amount of symmetries in the superconformal group gets its merit. Consistency with the Lorentz, conformal and $R$-symmetries puts strong restrictions on the possible terms in the supersymmetry variations. Keeping these restrictions in mind all the thinkable terms were added to the variation, with a priori unknown coefficients. These coefficients were then determined by consistency with the soft algebra. In extended supergravity one normally finds such soft algebras instead of conventional Lie algebras, see \cite{Batalin:1983jr,Henneaux:1990jq,Gomis:1995he} for a more detailed description of such algebras. In a soft algebra the commutator relations deform such that the parameters are dependent on the field content itself, for the $\N=3$ Weyl multiplet the soft algebra has been given in Section \ref{SSec:The nonlinear supersymmetry transformations}.
		
		Finally, we would like to elaborate on some possible future applications of the theory constructed in this paper. With the knowledge of the Weyl multiplet and its full supersymmetric variations one is able to research the possibilities for extending to higher order derivative theories as well. This was already done in \cite{Ciceri:2015qpa} for $\N=4$ conformal supergravity. However, because very little was known for the $\N=3$ case this is still an open problem.
		
		Also, applications in holography would be interesting with respect to this newly found theory. Recent papers, \cite{Karndumri:2016miq,Karndumri:2016tpf}, have discussed several of these applications concerning $\N=3$ Poincar\'e supergravity. It would be interesting to see in what way these results can be incorporated into the superconformal theory. In recent studies of double copy constructions the case of $\N=3$ supergravity has also been of interest \cite{Ferrara:2018iko}.

\medskip
\section*{Acknowledgments.}

\noindent We are grateful to A. Amariti, N. Bobev, L. Cassia, S. Ferrara, F. Gautason, S. Hegde, S. Penati, B. Sahoo, T. Van Riet and G. Tartaglino-Mazzucchelli for interesting and useful discussions.

The work is supported in part by the KU Leuven C1 grant ZKD1118 C16/16/005. JvM is a PhD Fellow of the Research Foundation - Flanders and his work is supported in part by the starting grant BOF/STG/14/032 from KU Leuven.

During the preparation of the second version of this paper, we became also aware of the work of \cite{Amariti:2017cyd}, where the superconformal ${\cal N}=3$ theory was used for constructing $D=2$ theories on twisted Riemann manifolds. These authors and also those of \cite{Hegde:2018mxv} pointed out numerical errors in the first version of this paper.
\newpage

\newpage
\appendix
		\section{Useful identities and conventions}\label{app: Usefull identities and calculations}
		\subsection{Conventions}
		In general the conventions of \cite{Freedman:2012zz} are used.  Here we will mention a few important notions that could potentially cause for confusion.
				Throughout this paper the mostly plus convention is used for the metric.
		When a two sided derivative is used it will be with a minus sign when the derivative works as a right acting operator:
		\begin{equation}
		A\stackrel{\leftrightarrow}{\partial}_{\mu} B := A\partial_{\mu} B - \partial_{\mu}(A)B.
		\end{equation}
		
A \textit{dualized} tensor is denoted with a tilde and we use the following definition of the dual:
		\begin{align}
		\tilde{G}_{\mu\nu} := -\frac{1}{2}\rmi e\,\levi{\mu\nu\rho\sigma}{}G^{\rho\sigma}.
		\end{align}
		
		\subsection{Chiral notations}\label{SSec:Chiral notations}
		In $\N$-extended supersymmetry the fields are, amongst others, representations of the $R$-symmetry group. Specifically for $\N=3$ supersymmetry the fields are representations of the group $\SU(3) \times \U(1)$. For the $\N=4$ case on the other hand we have that the $R$-symmetry is described by $\SU(4)$. Concretely this means that the fields in such theories have extra indices, besides the possible spacetime and spinor indices. These indices are denoted with Latin letters $i,j,k, \ldots$. Furthermore, we will use these indices to distinguish between fields and their charge conjugated versions. An example of this is given by the $R$-symmetry gauge field in the $\N=4$ and $3$ Weyl multiplets described in Sections \ref{Sec:The N=4 Weyl Multiplet} and \ref{Sec:Truncation to N=3} respectively. The complex conjugation of this field is given by
		\begin{equation}
		(\gav{\mu j}{\s \;i})^c = \gav{\mu \s i}{\s j}= -\gav{\mu i}{\s \;j},
		\end{equation}
where the latter equation is due to the antihermiticity of these fields.
		For spinors we can use these indices to denote their chirality. For instance, for the spinor $\Lambda_i $, which is in a vector representation of the $R$-symmetry group, we define
		\begin{equation}
		\Lambda_i = P_R \Lambda_i \Rightarrow (\Lambda_i)^c = \Lambda^i = P_L \Lambda^i.
		\end{equation}
		For different spinors we have used different conventions for their chirality. This should become clear if one keeps in mind that for the supersymmetry parameters we use the following conventions
		\begin{equation}
		\gammas{*}{} \susyparr{}{i} = \susyparr{}{i} \,,\qquad
		\gammas{*}{} \susyparrconformal{}{i} = -\susyparrconformal{}{i}.
		\end{equation}
		The chirality for all the other spinors then follow from consistencies in the supersymmetry transformations. For clarity all the conventions of the spinor chiralities with respect to the $\SU(\N)$ representation have been written in Table \ref{Tab:Spinor chiral notation}.
		\begin{table}
			\begin{center}
				\begin{tikzpicture}
				\clip node (m) [matrix,matrix of nodes,
				fill=black!20,inner sep=0pt,
				nodes in empty cells,
				nodes={minimum height=1.17cm,minimum width=2cm,anchor=center,outer sep=0,font=\fontsize{10}{12}\sffamily},
				row 1/.style={nodes={fill=black,text=white}},
				column 1/.style={text width=1.5cm,align=center,every even row/.style={nodes={fill=black!2.5}}},
				column 2/.style={text width=1.5cm,align=center,every even row/.style={nodes={fill=black!2.5}}},
				column 3/.style={text width=1.5cm,align=center,every even row/.style={nodes={fill=black!2.5}},},
				column 4/.style={text width=1.5cm,align=center,every even row/.style={nodes={fill=black!2.5}},},
				column 5/.style={text width=1.5cm,align=center,every even row/.style={nodes={fill=black!2.5}},},
				column 6/.style={text width=1.5cm,align=center,every even row/.style={nodes={fill=black!2.5}},},
				column 7/.style={text width=1.5cm,align=center,every even row/.style={nodes={fill=black!2.5}},}
				row 1 column 1/.style={nodes={fill=black, text=white}},
				prefix after command={[rounded corners=2mm]  (m.north east) rectangle (m.south west)}
				] {
					Theory $(\N)$ &	Spinor & Chirality & Spinor & Chirality \\
					3 &	$\curlL,\curlR $& L,R & $ \galL,\galR$ & L,R   \\
					4 &	$ \lambda_i $& R & $ \Lambda_i $ & L   \\
					3 &	$ \curxi{ij}{} $& R & $ \gaxi{ij}{} $ & L \\
					3 &	&  & $ \gaz{i}{} $ & R \\
					4 &	$ \xi_{k}^{ij} $& R & $\gaxi{k}{ij}$ & L\\
					3,4 &&  & $ \spgravitino{\mu}{i} $ & R \\
					3,4 &	$ \cursup{\mu}{i} $& R &$  \gravitino{\mu}{i} $ & L\\
					3,4 &	$Q_i$& L & $\susyparr{i}{}$& R \\
					3,4 &	$S_i$& R & $\susyparrconformal{i}{}$& L \\
				};
				\end{tikzpicture}
				\caption{\textit{The handedness of the different spinors with respect to the chiral notation.}}\label{Tab:Spinor chiral notation}
			\end{center}
		\end{table}
		
		\subsection{Traces and hermitian conjugation}
		In several supersymmetry variations we have used the notation
		\begin{equation}
		\delta V = W - \text{ trace },\quad \text{and}\quad \delta V = W + \hc.
		\end{equation}
		The meaning of these notations is the following. As a first example, for indices running over $i=1,\ldots ,4$:
\begin{equation}
  W^{[ij]}_k -\text{ trace }= W^{[ij]}_k +\frac{2}{3}\delta _k^{[i}W^{j]\ell}_\ell\,,
 \label{wijktraceless}
\end{equation}
where the coefficient of the second term is determined by the demand that $\delta ^k_i$ on this expression vanishes.
 Say we have a tensor $\gad{k\ell}{ij}$ in some representation of $\SU(\N)$ such that $\gad{k\ell}{ij} = \gad{[k\ell]}{[ij]} $ and $\gad{k\ell}{kj} = 0$. Then we will have to ensure that the supersymmetric variation of this tensor will have the same symmetry and trace properties. This is done by taking a variation of the following form
		\begin{equation}
		\delta \gad{k\ell}{ij} = V_{k\ell}^{ij} - \alpha \kron{[k}{[i} V^{j]p}_{\ell]p} - \beta \kron{k\ell}{ij} V_{pq}^{pq} \,,\qquad \kron{k\ell}{ij}=\delta ^{[k}_{[i}\delta ^{\ell]}_{j]}\,.
		\end{equation}
		The coefficients $\alpha$ and $\beta$ are then determined by making $\delta \gad{kj}{k\ell}$ vanish identically. In the case of $\SU(4)$, such that $i,j,\ldots=1,\ldots,4$, this is done by taking $\alpha = 2$ and $\beta  =  - \frac{1}{3}$. If the field $D$ is also hermitian we will have to ensure that the right-hand-side of the variation is also hermitian. The following example shows how this is done by adding the hermitian conjugated form of a variation. If we write that the tensor varies under a supersymmetry as
		\begin{equation}
		\delta \gad{k\ell}{ij} = \susyparrc{}{[i}W^{j]}_{[k\ell]} - \text{trace} + \hc\,.
		\end{equation}
and the complex conjugate of $W^i_{jk}=W^i_{[jk]}$ is $W_i^{jk}$ then this means
		\begin{equation}
		\delta \gad{k\ell}{ij} = \susyparrc{}{[i}W^{j]}_{[k\ell]}-\kron{[k}{[i}\susyparrc{}{j]}W^{p}_{\ell]p}+\bar \epsilon ^p \kron{[k}{[i}W^{j]}_{\ell]p}+\frac{1}{3}\kron{k\ell}{ij}\bar \epsilon ^pW_{pq}^q
+ \susyparrc{[k}{}W^{ij}_{\ell]}-\kron{[k}{[i}\susyparrc{\ell]}{}W_{p}^{j]p}+\bar \epsilon_p \kron{[k}{[i}W_{\ell]}^{j]p}+\frac{1}{3}\kron{k\ell}{ij}\bar \epsilon _pW^{pq}_q\,.
		\end{equation}
		
\subsection{Comparing the conventions to the literature}
\label{app:translate}
Much of the work presented in this paper is related to results already found in \cite{Bergshoeff:1980is}. In this appendix we will clarify the differences in terms of the conventions between this work and theirs.

For the (anti-)symmetrization we have taken the "weight one" convention such that
\begin{align}
V_{(i}W_{j)} =\frac12 \left(V_i W_j + V_j W_i\right), \quad \text{and similar for the antisymmetrization},
\end{align}
in contrast with \cite{Bergshoeff:1980is} where no factor $1/2$ is taken. Another difference is that the Levi-Civita symbol with space-time indices, $\varepsilon_{abcd}$, in \cite{Bergshoeff:1980is} should be replaced with $-\rmi\varepsilon_{abcd}$ to compare the results. The antisymmetric gamma matrix with two indices in \cite{Bergshoeff:1980is} is defined as
\begin{align}
\sigma_{ab} = \frac12 \gamma_{ab} .
\end{align}
Furthermore, for the gauge fields some conventions have changed (the left-hand side are symbols used in \cite{Bergshoeff:1980is}, while the right-hand side consists of symbols used in this work)
\begin{equation}
\begin{aligned}
V_{\mu \phantom{i}j}^{\phantom{\mu}i} \Leftarrow& V_{\mu j}^{\phantom{\mu j} i},\qquad \phi_{\mu i} \Leftarrow 2 \phi_{\mu i},\\
f_{\mu}^a \Leftarrow& 2 f_\mu^a, \hspace{1.1 cm} a_\mu \Leftarrow -\rmi \Upsilon_\mu
\end{aligned}
\end{equation}
The final differences are in the definitions of the parameters of the soft algebra
\begin{equation}
\begin{aligned}
\epsilon_i \Leftarrow \frac12 \epsilon_i,\qquad \epsilon^{ab} \Leftarrow - \lambda_{ab},\\
\Lambda_K^a \Leftarrow 2 \lambda_K^a, \qquad \Lambda \Leftarrow -\lambda_T.
\end{aligned}
\end{equation}
\newpage
\providecommand{\href}[2]{#2}\begingroup\raggedright\endgroup

\end{document}